\documentclass[preprint,nofootinbib,superscriptaddress,noshowkeys,showpacs]{revtex4}
\usepackage{bm}
\usepackage{amsmath}
\usepackage{amsfonts}
\usepackage[dvips]{graphicx}

\usepackage{ulem}
\usepackage{color}

\begin{document}
\title{Signature of strange dibaryons in kaon- and photon-induced reactions}
\author{Shota Ohnishi}
\email{s_ohnishi@riken.jp}
\affiliation{Department of Physics, Tokyo Institute of Technology, Tokyo 152-8551, Japan}
\affiliation{RIKEN Nishina Center, Wako, Saitama 351-0198, Japan}

\author{Yoichi Ikeda}
\affiliation{RIKEN Nishina Center, Wako, Saitama 351-0198, Japan}

\author{Hiroyuki Kamano}
\affiliation{Research Center for Nuclear Physics, Osaka University, Osaka 567-0047, Japan}

\author{Toru Sato}
\affiliation{Department of Physics, Osaka University, Osaka 560-0043, Japan}
\affiliation{J-PARC Branch, KEK Theory Center, Institute of Particle and Nuclear Studies, KEK, 
203-1, Shirakata, Tokai, Ibaraki, 319-1106, Japan}
\date{\today}

\begin{abstract}
We examine how the signature of the strange-dibaryon resonances
with $I=1/2$ and $J^\pi =0^-$ shows up in 
scattering amplitudes and observables 
of the three-body $\bar{K}NN$-$\pi Y N~(Y=\Sigma,~\Lambda)$ system 
on the physical real energy axis.
The so-called point method is applied to handle logarithmic singularities
that appear in solving the Alt-Grassberger-Sandhas equations for 
the real scattering energies.
By taking two different kinds of models for the two-body 
$\bar{K}N$-$\pi \Sigma$ subsystem, 
both of which reproduce the available data 
equally well but give quite a different resonance-pole structure 
for $\Lambda(1405)$,
we also investigate whether the strange-dibaryon production reactions 
can be used for disentangling the nature of $\Lambda(1405)$.
\end{abstract}
\maketitle

\section{Introduction}
In recent years, the strange dibaryons with $I=1/2$ and $J^\pi =0^-$
have been studied actively
as the simplest kaonic nuclei~\cite{Akaishi:2002bg} 
in the three-body $\bar{K}NN$-$\pi Y N$ system.
A number of theoretical studies to search for the strange dibaryons 
have been performed with the variational method
\cite{Yamazaki:2002uh,Wycech:2008wf,Dote:2008in,Barnea:2012qa} and
the Alt-Grassberger-Sandhas (AGS) equations
\cite{Shevchenko:2006xy,Ikeda:2007nz,Ikeda:2010tk},
employing the phenomenological potentials 
\cite{Shevchenko:2006xy,Yamazaki:2002uh,Wycech:2008wf} or the effective chiral Lagrangian
\cite{Ikeda:2007nz,Ikeda:2010tk,Dote:2008in,Hyodo:2007jq,Barnea:2012qa}
for the meson-baryon and baryon-baryon interactions.
All the studies support the existence of the strange dibaryons
as resonance states in the energy region between the $\bar{K}NN$ and
$\pi \Sigma N$ thresholds.
However, the resonance energies predicted in those studies are still 
highly model dependent. 
For example,
the models with energy-independent potentials~\cite{Yamazaki:2002uh,Wycech:2008wf,Shevchenko:2006xy,Ikeda:2007nz} 
give resonance energies 
lower than those with energy-dependent potentials~\cite{Dote:2008in,Barnea:2012qa,Ikeda:2010tk}.

In parallel with the theoretical works mentioned above,
experimental searches for the strange dibaryons have also been done
by the FINUDA Collaboration \cite{Agnello:2005qj}, the OBELIX
Collaboration \cite{Bendiscioli:2007zza}, and the DISTO
Collaboration \cite{Yamazaki:2008hm}. 
Further data will become available
from SPring-8 (LEPS Collaboration~\cite{Parker:2008zz}) and GSI (FOPI
Collaboration~\cite{Suzuki:2010zzg}),
and new experiments are planned at 
J-PARC (E15~\cite{Iwasaki} and E27~\cite{Nagae} experiments) and DA$\Phi$NE (AMADEUS
Collaboration~\cite{Zmeskal:2011zz}).

In our previous works~\cite{Ikeda:2007nz,Ikeda:2010tk}, 
we have investigated a possible existence of
the strange-dibaryon resonances in the three-body $\bar{K}NN$-$\pi \Sigma N$ system.
This has been achieved by searching for resonance poles
of the three-body amplitudes in the complex energy plane, 
where the amplitudes are obtained by solving 
the coupled-channel AGS equations.
There, two models, the energy-independent (E-indep) and
the energy-dependent (E-dep) models, have been employed 
for the $s$-wave meson-baryon interactions, both of which are derived from 
the leading-order term of the effective chiral Lagrangian 
but those have different off-shell behavior.
As a result, we have found one resonance pole of 
the strange dibaryon for the E-indep model
and two for the E-dep model, which are summarized in Table~\ref{res_ene}.
This result indicates that off-shell behavior of the meson-baryon interactions of 
the two-body $\bar K N$-$\pi Y$ subsystem
is crucial for the resulting pole positions of the strange-dibaryon resonances.

\begin{figure*}[thb]
\includegraphics[width=\textwidth,clip]{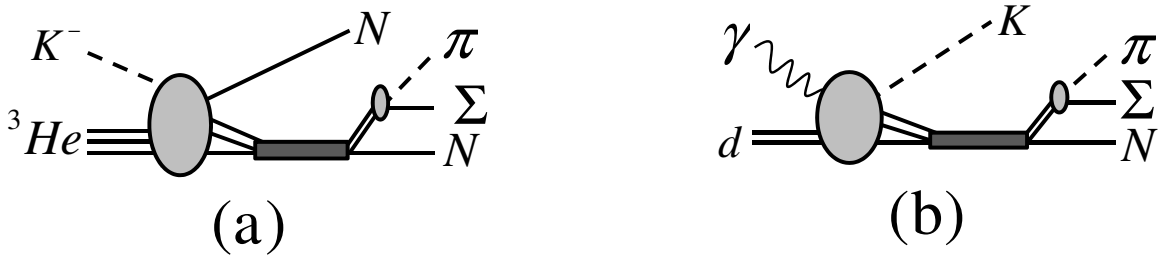}
\caption{Examples of the typical (a)kaon- and (b)photon-induced strange-dibaryon production reactions.
The strange-dibaryon resonances would be produced
in thick shaded boxes.}
\label{reaction}
\end{figure*}

\begin{table}[thb]
\caption{
Pole masses $M_R$ of the strange-dibaryon resonances obtained 
in our previous works~\cite{Ikeda:2007nz,Ikeda:2010tk}.
See the text for the explanation on the E-indep and E-dep models.
The $\bar K NN$ and $\pi\Sigma N$ threshold energies are 2370 and 2267 MeV,
respectively.
}
\label{res_ene}      
\begin{ruledtabular}
\begin{tabular}{lccc}
 & $\text{Re}(M_R)$ (MeV) & $-\text{Im}(M_R)$ (MeV) \\
\hline
E-indep model & 2312-2326 & 17-20\\
E-dep model & 2354-2361 & 17-23 \\
 & 2281-2303 & 122-160
\end{tabular}
\end{ruledtabular}
\end{table}

Most of the theoretical studies have presented only pole positions of 
the strange-dibaryon resonances.
However, those are not a quantity that can be directly measured in experiments.
To examine the existence of the strange dibaryons in connection with experiments,
one has to compute the cross sections of strange-dibaryon production reactions
consistently in the same framework.
The strange-dibaryon resonances can be produced via, for example,
kaon- and photon-induced reactions on light
nuclei such as $^3$He and deuterons (Fig.~\ref{reaction}).
Then the signal of the resonances would be observed in the invariant-mass and/or
missing-mass distributions of the decay products.
A couple of such studies have been performed by Koike-Harada~\cite{Koike:2009mx}
and Yamagata-Sekihara {\it et al.}~\cite{YamagataSekihara:2008ji} 
on the basis of the optical potential approach.

In this work, we examine how the signature of the strange dibaryons 
shows up in the observables of the three-body reactions
by applying our approach based on the coupled-channel AGS equations 
developed in Refs.~\cite{Ikeda:2007nz,Ikeda:2010tk}.
It is well known that logarithmic singularities appear
when one solves the AGS equations for the breakup reactions at real scattering energies.
We handle those singularities numerically by making use of the so-called point-method 
proposed by Schlessinger~\cite{Schlessinger:1968} and developed by 
Kamada~\textit{et al.}~\cite{Kamada:2003xy}.
With this method, we examine the behavior
of the quasi-two-body amplitudes (the thick shaded boxes in Fig. \ref{reaction})
of the $\bar{K}NN$-$\pi Y N$ system 
at real scattering energies between
the $\bar{K}NN$ and $\pi \Sigma N$ thresholds.
As a first step toward developing a model to compute reaction cross sections 
measured at facilities such as J-PARC and SPring-8 
(e.g., the reactions in Fig.~\ref{reaction}), 
we examine the ``transition probability'' of a strange-dibaryon 
production reaction, $(Y_K)_{I=0} + N \to \pi + \Sigma + N$, where $(Y_K)_{I=0}$ is an ``isobar''
of $\bar K N$ states with isospin $I=0$.
We also give an estimation of the probability for the kaon absorption process
$(Y_K)_{I=0} + N \to \Lambda + N$.

In Sec.~\ref{sec:three-body}, we explain the AGS equations for the three-body
$\bar{K}NN$-$\pi YN(Y=\Sigma,\Lambda)$ system and 
present the transition probability formula for break-up reactions.
Then, we present the two-body meson-baryon
interactions used in this work in Sec.~\ref{sec:two-body}.
The computed quasi-two-body amplitudes as well as transition probabilities
for $(Y_K)_{I=0} + N \to \pi + \Sigma + N$ are presented in Sec.~\ref{sec:result}.
The summary is given in Sec.~\ref{sec:summary}.
A brief description of the point method is presented in the Appendix.
\section{Three-body Equations}
\label{sec:three-body}

\subsection{Alt-Grassberger-Sandhas equations}
\label{sec:ags}

Throughout this paper, we assume that the three-body processes take place via
separable two-body interactions, which have the following form 
in the two-body center-of-mass (c.m.) frame,
\begin{align}
V_{(\alpha)_I i,(\beta)_I i}(\vec q_i ~',\vec q_i ; E) = 
g_{(\alpha)_I }^\ast(\vec q_i ~') \lambda_{(\alpha)_I i,(\beta)_I i}(E)
 g_{(\beta)_I }(\vec q_i) ~,
\label{eq:v_sepa}
\end{align}
where $g_{(\alpha)_I}(\vec q_i)$ is the cutoff factor
of the two-body channel $\alpha(=jk)$, with relative momentum $\vec q_i$ and
isospin $I$, and
$E$ is the total energy of the two-body system.
In the three-body system, we define the two-body energy $E$ as
$E=W-E_i(\vec p_i)$, with the three-body energy $W$ and the spectator
particle energy $E_i(\vec p_i)$, where
$\vec p_i$ is the relative momentum of the spectator particle $i$.
The explicit forms of each two-body interaction are presented in detail in Sec.~\ref{sec:two-body}.

\begin{table*}[thb]
\caption{
Indices specifying the ``isobars.''
}
\label{isobar}      
 \begin{ruledtabular}
\begin{tabular}{lcccccc}
Isobar& Allowed isospin(s) & Spectator particle & three-body Fock space\\
\hline
$(Y_K)=({\bar K N_2}),(\bar{K}N_1)$ & 0, 1  & $N_1,N_2$ &$\left|
		 N_1N_2\bar{K}\right\rangle$\\
$(Y_\pi)=(\pi\Sigma)$ & 0, 1 & $N$
	     &$\left|N\Sigma\pi\right\rangle,\left|\Sigma N\pi\right\rangle$\\
$(Y_\pi)=(\pi\Lambda)$ & 1 & $N$
	     &$\left|N\Lambda\pi\right\rangle,\left|\Lambda N\pi\right\rangle$\\
$(d)=(NN)$ & 1 & $\bar K$ &$\left| N_1N_2\bar{K}\right\rangle$\\
$(N^*)=(\pi N)$  & 1/2, 3/2 & $\Sigma$ &$\left| \Sigma N\pi\right\rangle,\left|N\Sigma\pi\right\rangle$\\
$(N^*)=(\pi N)$ & 1/2 & $\Lambda$ &$\left| \Lambda N\pi\right\rangle,\left|N \Lambda\pi\right\rangle$\\
$(d_y)=(\Sigma N)$ & 1/2, 3/2 & $\pi$ &$\left| \Sigma N\pi\right\rangle,\left|N\Sigma\pi\right\rangle$\\
$(d_y)=(\Lambda N)$ & 1/2 & $\pi$&$\left| \Lambda N\pi\right\rangle,\left|N\Lambda\pi\right\rangle$
\end{tabular}
 \end{ruledtabular}
\end{table*}

The assumption above implies that two-body subsystems in the three-body processes form 
an ``isobar'' and thus the processes can be described as a quasi-two-body scattering of 
the isobar and the spectator particle.
The quasi-two-body amplitudes, $X_{(\alpha)_Ii,(\beta)_{I'}j}(\vec p_i, \vec p_j; W)$, 
are then obtained by solving
the AGS equations \cite{1963PhRv..132..485A,Alt:1967fx},
\begin{align}
	X_{(\alpha)_Ii,(\beta)_{I'}j}({\vec p}_i,{\vec p}_j,W)&=(1-\delta_{ij})
	Z_{(\alpha)_Ii,(\beta)_{I'}j}({\vec p}_i,{\vec p}_j,W)
\nonumber\\
&+\sum_{(\gamma),(\delta)}\sum_{I''}\sum_{n\ne i}\int d{\vec p}_n Z_{(\alpha)_Ii,(\gamma)_{I''}n}({\vec p}_i,{\vec p}_n,W)
\nonumber\\
&\qquad\times
	\tau_{(\gamma)_{I''}n,(\delta)_{I''}n}\left(W-E_n(\vec p_n),\vec p_n\right) X_{(\delta)_{I''}n,(\beta)_{I'}j}
({\vec p}_n,{\vec p}_j,W)~.
	\label{AGS}
\end{align}
Here $(\alpha)_I$ denotes the isobar formed by a two-particle pair
$\alpha$ with isospin $I$;
the subscripts $i,j,$ and $n$ represent the spectator particles.
The notations for the isobars are summarized in Table~\ref{isobar}.
As is shown in Sec.~\ref{sec:two-body}, in this work 
we include only the $^1S_0$ partial wave for the $NN$ interaction,
and thus only the isospin $I=1$ state is allowed for the isobar $(d)$.

\begin{figure*}
 \includegraphics[width=\textwidth,clip]{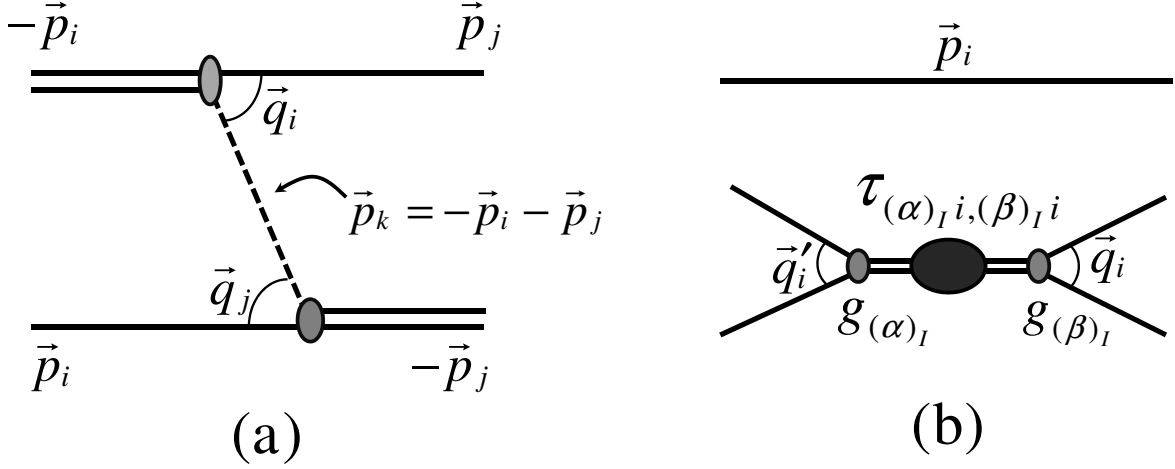}
 \caption{(a) One-particle exchange interaction $Z_{(\alpha)_Ii,(\beta)_{I'}j}({ p}_i,{
 p}_j,W)$. (b) Isobar propagator $\tau_{(\alpha)_Ii,(\beta)_{I}i}(W-E_i(\vec p_i),\vec p_i)$.}
 \label{z-diagram}
\end{figure*}

The driving term $Z_{(\alpha)_Ii,(\beta)_{I'}j}({\vec p}_i,{\vec p}_j;W)$ describes a particle-exchange potential 
given by [see Fig.~\ref{z-diagram}(a) for the kinematics]
\begin{align}
Z_{(\alpha)_Ii,(\beta)_{I'}j}({\vec p}_i,{\vec p}_j;W) =
\frac{g_{(\alpha)_I }(\vec q_i)g_{(\beta)_{I'} }^\ast(\vec q_j)}{W - E_i(\vec p_i) - E_j(\vec p_j) - E_k (\vec p_k) + i\epsilon},
\label{eq:z-diagram}
\end{align}
where $E_i(\vec p_i)$ and $E_j(\vec p_j)$ are the energies of the
spectator particles $i$ and $j$, respectively;
$E_k(\vec p_k)$ with $\vec p_k = - \vec p_i - \vec p_j$ is the energy of the exchange particle $k$;
and $\vec q_i$ ($\vec q_j$) is the relative momentum between the exchange-particle 
and the spectator-particle $j$ ($i$).
In the nonrelativistic kinematics, we have 
$E_{n}(\vec p_n) = m_n + \vec p_{n}^2/(2m_{n})$ ($n=i,j,k$)
and $\vec q_{i,j} = (m_{k,i} \vec p_{j,k} - m_{j,k}\vec p_{k,i})/(m_{j,k}+m_{k,i})$.
The $s$-wave projection of $Z_{(\alpha)_Ii,(\beta)_{I'}j}(\vec p_i, \vec p_j; W)$ is given by
\begin{equation}
Z_{(\alpha)_Ii,(\beta)_{I'}j}(p_i, p_j; W) = \frac{1}{2}\int_{-1}^1 d(\cos\theta) Z_{(\alpha)_Ii,(\beta)_{I'}j}(\vec p_i, \vec p_j; W),
\label{eq:z-diagram-s}
\end{equation}
with $\cos\theta = \hat p_i \cdot \hat p_j$.

The isobar propagator, $\tau _{(\alpha)_Ii,(\beta)_{I}i}\left(W-E_i(\vec p_i),\vec p_i\right)$ as illustrated in 
Fig.~\ref{z-diagram}(b), is given in the nonrelativistic kinematics by solving
the following Lippmann-Schwinger equations:
\begin{align}
 \tau_{(\alpha)_Ii,(\beta)_{I}i}(W-E_i(\vec p_i),\vec p_i)  &= 
  \lambda_{(\alpha)_I i,(\beta)_{I} i}\nonumber\\
  &+\sum_{(\gamma)}\int q_i^2d{ q}_i
  \frac{\lambda_{(\alpha)_I i,(\gamma)_I i}|g_{(\gamma)_I}({ q}_i)|^2}
  {W-E_i({\vec p}_i)-E_{jk}({\vec p}_i,{\vec q}_i)}\tau_{(\gamma)_Ii,(\beta)_{I}i}(W-E_i(\vec p_i),\vec p_i)~~.
\end{align}
Here, $E_{jk}({\vec p}_i,{\vec q}_i)$ is the energy of the interacting pair ($jk$),
$
E_{jk}({\vec p}_i,{\vec q}_i) = m_j+m_k+{\vec p}~_i^2/2(m_j+m_k)+{\vec
q}~_i^2/2\mu_i
$
with the reduced mass defined as $\mu_i =m_j m_k/(m_j +m_k)$.

After taking antisymmetrization for the two-nucleon states in the three-body processes, 
the AGS equations (\ref{AGS}) are formally written as
(suppressing all indices other than those of the isobars)
\begin{align}
 &\begin{pmatrix}
   X_{(Y_K),(Y_K)}\\
   X_{(Y_\pi), (Y_K)}\\
   X_{(d),(Y_K)}\\
   X_{(N^*), (Y_K)}\\
   X_{(d_y), (Y_K)}
  \end{pmatrix}
 =
 \begin{pmatrix}
  Z_{(Y_K), (Y_K)}\\
  0\\
  Z_{(d), (Y_K)}\\
  0\\
  0
 \end{pmatrix}
\nonumber\\
 &-
  \begin{pmatrix}
  Z_{(Y_K),(Y_K)}\tau_{(Y_K),(Y_K)}&Z_{(Y_K),(Y_K)}\tau_{(Y_K),(Y_\pi)}&2Z_{(Y_K),(d)}\tau_{(d),(d)}&0&0\\
  0&0&0&Z_{(Y_\pi), (N^*)}\tau_{(N^*),(N^*)}&Z_{(Y_\pi), (d_y)}\tau_{(d_y),(d_y)}\\
  Z_{(d),(Y_K)}\tau_{(Y_K),(Y_K)}&Z_{(d),(Y_K)}\tau_{(Y_K),(Y_\pi)}&0&0&0\\
  Z_{(N^*), (Y_\pi)}\tau_{(Y_\pi), (Y_K)}&Z_{(N^*), (Y_\pi)}\tau_{(Y_\pi), (Y_\pi)}&0&0&Z_{(N^*), (d_y)}\tau_{(d_y),(d_y)}\\
  Z_{(d_y), (Y_\pi)}\tau_{(Y_\pi), (Y_K)}&Z_{(d_y), (Y_\pi)}\tau_{(Y_\pi), (Y_\pi)}&0&Z_{(d_y), (N^*)}\tau_{(N^*), (N^*)}&0
\end{pmatrix}
 \nonumber\\
 &\times
 \begin{pmatrix}
   X_{(Y_K),(Y_K)}\\
   X_{(Y_\pi), (Y_K)}\\
   X_{(d) ,(Y_K)}\\
   X_{(N^*), (Y_K)}\\
   X_{(d_y), (Y_K)}
 \end{pmatrix}.\label{coupled-AGS}
\end{align}

\subsection{Break-up reactions}

In this subsection, we present formulas for computing transition probability
of the quasi-two-body to three-body reaction, $(Y_K)_{I=0} + N \rightarrow \pi + \Sigma +N$.
For this purpose, we first need to define the amplitudes
of the $(Y_K)_{I=0} + N \rightarrow \pi + \Sigma +N$ reaction. 
This is because within our formulation
the well-defined amplitudes are of the three-body to three-body scatterings,
where all the external particles are stable against strong interactions.
The relevant amplitude here is of the
$(\bar K + N) + N \to (Y_K)_{I=0} + N \rightarrow \pi + \Sigma +N$ reaction,
which is given in a concise notation as
\begin{eqnarray}
T_{\pi\Sigma N\leftarrow (\bar K N) N} = 
\sum_{(\alpha) i = \pi\Sigma N} \sum_{(\gamma)}\sum_I
g^\ast_{(\alpha)_I} \tau_{(\alpha)_Ii,(\gamma)_Ii} X_{(\gamma)_Ii,(Y_K)_{I=0} N}
\tau_{(Y_K)_{I=0} N,(Y_K)_{I=0} N} g_{(Y_K)_{I=0}} ,
\end{eqnarray}
where the summation of $(\alpha) i$ is taken for all possible combinations of $\pi\Sigma N$.
Now let us consider the isobar $(Y_K)_{I=0}$ as an actual resonance state of the two-body reactions. 
(Note that we originally introduced notion of the isobars 
just for the sake of convenience in our formulation
and did not take them as actual resonances.)
Near a resonance pole of the isobar propagator $\tau_{(Y_K)_{I=0},(Y_K)_{I=0}}(E)$,
the two-body amplitude for $\bar K N _{I=0}\to \bar K N_{I=0}$ can be approximated as
\begin{eqnarray}
t_{(Y_K)_{I=0},(Y_K)_{I=0}}(E) & = & g^*_{(Y_K)_{I=0}} \tau_{(Y_K)_{I=0},(Y_K)_{I=0}}(E,\vec 0)g_{(Y_K)_{I=0}} \nonumber\\
    & \sim & g^*_{(Y_K)_{I=0}} \frac{\sqrt{R_{(Y_K)_{I=0}}}\sqrt{R_{(Y_K)_{I=0}}}}{E - M + i\Gamma/2}g_{(Y_K)_{I=0}}
\nonumber\\
    & \equiv & \bar g^*_{\bar K N \leftarrow Y_K} \frac{1}{E - M + i\Gamma/2}
\bar g_{Y_K \leftarrow \bar K N},
\end{eqnarray}
where $M-i\Gamma/2$ is the resonance pole position of $\tau_{(Y_K)_{I=0},(Y_K)_{I=0}}(E)$ and 
$R_{(Y_K)_{I=0}}$ is the residue of $\tau_{(Y_K)_{I=0},(Y_K)_{I=0}}$ at the pole. 
Also,
$\bar{g}_{Y_K \leftarrow \bar K N} = \sqrt{R_{(Y_K)_{I=0}}} g_{(Y_K)_{I=0}}$  
[$\bar{g}_{\bar K N \leftarrow Y_K} = \sqrt{R_{(Y_K)_{I=0}}} g^*_{(Y_K)_{I=0}}$]
can be interpreted as a vertex function for the process 
$\bar K N _{I=0}\to (Y_K)_{I=0}$ [$(Y_K)_{I=0} \to \bar K N_{I=0}$].
Within this approximation, the three-body amplitude can be written as
\begin{eqnarray}
T_{\pi\Sigma N\leftarrow (\bar K N) N} &=& 
\sum_{(\alpha) i = \pi\Sigma N} \sum_{(\gamma)}\sum_I
g^\ast_{(\alpha)_I } \tau_{(\alpha)_Ii,(\gamma)_Ii} X_{(\gamma)_Ii,{(Y_K)}_{I=0} N} 
\nonumber\\
&\times&\tau_{{(Y_K)}_{I=0}N,{(Y_K)}_{I=0}N}(W-E_N(\vec p_N),\vec p_N) g_{(Y_K)_{I=0}} 
\nonumber\\
&\sim &
\sum_{(\alpha) i = \pi\Sigma N} \sum_{(\gamma)}\sum_I
g^\ast_{(\alpha)_I} \tau_{(\alpha)_Ii,(\gamma)i} X_{(\gamma)i,{(Y_K)}_{I=0} N}\nonumber\\
&\times&\ \sqrt{R_{{(Y_K)}_{I=0}}}
G_{{(Y_K)}_{I=0}}(W-E_N(\vec p_N),\vec p_N) \bar g_{Y_K\leftarrow \bar K N} ,
\label{eq:three-aprx}
\end{eqnarray}
where $G_{{(Y_K)}_{I=0}}(W-E_N(\vec p_N),\vec p_N)$ is the $(Y_K)_{I=0}$ resonance propagator
in the existence of a spectator nucleon with momentum $\vec p_N$.
From Eq.~(\ref{eq:three-aprx}), it is reasonable to define the $T$ matrix of 
$(Y_K)_{I=0} + N \rightarrow \pi + \Sigma +N$ as
\begin{eqnarray}
T_{\pi\Sigma N\leftarrow (Y_K)_{I=0} N} & = &
\sum_{(\alpha) i = \pi\Sigma N} \sum_{(\gamma)}\sum_I
g^\ast_{(\alpha)_I } \tau_{(\alpha)_Ii,(\gamma)_Ii} X_{(\gamma)_Ii,(Y_K)_{I=0} N} \sqrt{R_{(Y_K)_{I=0}}} .
\end{eqnarray}

The $s$-wave projection of the scattering amplitudes for the
$(Y_K)_{I=0}+N\rightarrow \pi +\Sigma+ N$ reaction
are then given by 
\begin{align}
&T_{\pi\Sigma N\text{-}(Y_K)_{I=0}N}(\vec q_N,\vec p_N, p'_N,W)
\nonumber\\
&=(4\pi)^{-3/2}\sum_I
\nonumber\\
& \times\{ 
\left|[[\pi\otimes \Sigma]_{(Y_\pi)_I}\otimes N]_{\Gamma}\right\rangle
g_{(Y_\pi)_I }(q_N) \tau_{(Y_\pi)_I N, (Y_K)_{I}N}\left(W-E_N(\vec p_N),\vec p_N\right) X_{(Y_K)_{I}N,(Y_K)_{I=0}N}(p_N, p_N',W)
\nonumber\\
&+\left|[[\pi\otimes \Sigma]_{(Y_\pi)_I}\otimes N]_{\Gamma}\right\rangle
 g_{(Y_\pi)_I }(q_N) \tau_{(Y_\pi)_I N,(Y_\pi)_I N}\left(W-E_N(\vec p_N),\vec p_N\right) X_{(Y_\pi)_I N, (Y_K)_{I=0}N}( p_N, p_N',W)
\nonumber\\
&+\left|[[\pi\otimes N]_{(N^*)_I}\otimes \Sigma]_\Gamma \right\rangle
g_{(N^*)_I}(q_\Sigma) \tau_{(N^*)_I\Sigma, (N^*)_I\Sigma}\left(W-E_\Sigma(\vec p_\Sigma),\vec p_\Sigma\right) X_{(N^*)_I \bar K, (Y_K)_{I=0} N}( p_\Sigma, p_N',W)
\nonumber\\
&+\left|[[\Sigma\otimes N]_{(d_y)_I}\otimes\pi]_\Gamma \right\rangle
g_{(d_y)_I}(q_\pi) \tau_{(d_y)_I\pi, (d_y)_I\pi}\left(W-E_\pi(\vec p_\pi),\vec p_\pi\right) X_{(d_y)_I \pi, (Y_K)_{I=0} N}( p_\pi, p_N',W)
\}
\nonumber\\
& \times \left\langle[(Y_K)_{I=0}\otimes N']_{\Gamma'}\right|\sqrt{R_{(Y_K)_{I=0}}}~,
\label{eq:t_break}
\end{align}
where $\left|[A\otimes B]_a \otimes C]_b\right\rangle$, with $(ABC) =(\pi\Sigma N)$, 
and $\left|[(Y_K)_{I=0}\otimes N]_b\right\rangle$
are the spin-isospin wave functions of the final and initial
states, and
$X_{(\alpha)_Ii,(\beta)_{I'}j}( p, p',W)$ is the $s$-wave projection of the quasi-two-body amplitudes 
given in Eqs.~(\ref{AGS}) and~(\ref{coupled-AGS}).
The momenta $ q_\Sigma$, $ q_\pi$, $ p_\Sigma$ and $ p_\pi$ are functions
of $\vec q_N$ and $\vec p_N$, i.e., 
$ q_\Sigma(\vec q_N,\vec p_N)$, $ q_\pi(\vec q_N,\vec p_N)$, $ p_\Sigma(\vec q_N,\vec p_N)$, 
and $ p_\pi(\vec q_N,\vec p_N)$.

Using Eq.~(\ref{eq:t_break}), 
we define the transition probability
of $(Y_K)_{I=0}+N\rightarrow \pi +\Sigma+ N$ as follows,
\begin{align}
 w( p'_N,W)=2\pi\int d^3\vec p_Nd^3\vec
 q_N\sum_{f\bar{i}}\delta\left(W-M-\frac{\vec p^2_N}{2\eta_N}-\frac{\vec
 q^2_N}{2\mu_N}\right)\left|T_{\pi\Sigma N-(Y_K)_{I=0}N}\left(\vec q_N,\vec
 p_N, p'_N,W\right)\right|^2~.
 \label{eq:cross}
\end{align}
\subsection{Kaon absorption reaction}
The two-body $\Lambda N$ channel is
one of the important decay channels of the strange dibaryons.
The main process of such a two-body decay
is expected to be the successive process with the kaon absorption, i.e.,
$\text{``strange dibaryon''}\to \bar{K} + N + N \to \Lambda + N$.
Therefore, we also evaluate the transition probability for the
$(Y_K)_{I=0} + N \to \bar{K} + N + N \to \Lambda + N$ reaction,
so that we can examine how differently the contribution of the strange dibaryons 
emerges to the absorption and breakup reaction cross sections.
For this purpose, we start with the three-body scattering amplitude of the
$(Y_K)_{I=0} + N \rightarrow \bar{K} + N + N$ reaction,
\begin{align}
&  T_{\bar{K} N N\text{-}(Y_K)_{I=0}N}(\vec q_N, \vec p_N, p'_N,W)
\nonumber\\
&= (4\pi)^{-3/2}\sum_{I=0,1}
\left|[[\bar{K} \otimes N]_{(Y_K)_I}\otimes N]_{\Gamma}\right\rangle
g_{(Y_K)_I }(q_N) \tau_{(Y_K)_I N, (Y_K)_{I}N}\left(W-E_N(\vec p_N),\vec p_N\right) 
\nonumber\\
& \times  X_{(Y_K)_{I}N,(Y_K)_{I=0}N}(p_N, p_N',W)
\left\langle[(Y_K)_{I=0}\otimes
 N']_{\Gamma'}\right|\sqrt{R_{(Y_K)_{I=0}}}~,
\label{eq:knn-knn}
\end{align}
where we follow the same convention as in Eq.~(\ref{eq:t_break}).
The transition probability of the kaon absorption reaction $w_{\rm abs}(p'_N, W)$
is then given by
\begin{align}
 w_{\rm abs}( p'_N,W)=
 2 \pi \int d^3\vec p_{\Lambda}
 \sum_{f\bar{i}} 
 \delta \left(W-(M_N+M_{\Lambda})-\frac{\vec p^2_{\Lambda}}{2\mu_{\Lambda N}}\right)
 \left|T_{\Lambda N-(Y_K)_{I=0}N}\left(\vec p_{\Lambda}, p'_N,W\right)\right|^2~,
 \label{eq:w_abs}
\end{align}
with
\begin{align}
 & T_{\Lambda N\text{-}(Y_K)_{I=0}N}\left(\vec p_{\Lambda}, p'_N,W\right) \nonumber \\
 & =
 \int d^3 \vec p_N
 V_{\rm abs}(\vec p_{\bar{K}}, \vec p_N) 
 \frac{1}{W - E_{\bar{K}}(\vec p_{\bar{K}}) - E_{N}(\vec p_N) - E_{N}(-\vec p_{\Lambda})}
 T_{\bar{K} N N\text{-}(Y_K)_{I=0}N}(\vec q_N, \vec p_N, p'_N,W) ~~.
\label{eq:knn-ln}
\end{align}
Here $\mu_{\Lambda N}$ and $\vec p_{\Lambda} = \vec p_{\bar{K}} + \vec p_N$
denote the reduced mass of $\Lambda N$ and the momentum of the $\Lambda$ particle 
in the final state, respectively, and
$V_{\rm abs}$ represents the kaon absorption vertex whose explicit expression
is given in Sec~\ref{sec:abs-int}.

\section{Model of Two-body Interactions}
\label{sec:two-body}

Now we present explicit forms of the two-body interactions [Eq.~(\ref{eq:v_sepa})] 
used in this work.
We first consider the meson-baryon interactions (Sec.~\ref{sec:mb-int}) and then
consider the baryon-baryon interactions (Sec.~\ref{sec:bb-int}).
In this section, we suppress indices of the spectator.

\subsection{Meson-baryon interaction}
\label{sec:mb-int}

As done in our earlier works~\cite{Ikeda:2007nz,Ikeda:2010tk}, 
we consider two kinds of models for the $s$-wave meson-baryon interactions,
which are called the E-indep and E-dep models, respectively.
The explicit forms are given by
\begin{equation}
V^{\text{E-indep}}_{(\alpha)_I (\beta)_{I}}(q',q)
=-C_{(\alpha)_I(\beta)_{I}}\frac{1}{32\pi^2 F_\pi^2}
\frac{m_\alpha +m_\beta}{\sqrt{\mathstrut m_\alpha m_\beta}}
g_{(\alpha)_I}(q')
g_{(\beta)_I}(q),
\label{eq:e-indep}
\end{equation}
for the E-indep model, and by
\begin{equation}
V^{\text{E-dep}}_{(\alpha)_I {(\beta)_I}}(q',q;E)
=-C_{(\alpha)_I{(\beta)_I}}\frac{1}{32\pi^2 F_\pi^2}
\frac{2E-M_\alpha -M_\beta}{\sqrt{\mathstrut m_\alpha m_\beta}}
g_{(\alpha)_I}(q')
g_{(\beta)_I}(q),
\label{eq:e-dep}
\end{equation}
for the E-dep model.
Here, $m_\alpha$ ($M_\alpha$) is the meson (baryon) mass of the channel $\alpha$; 
$q'$ ($q$) is the magnitude of relative momentum of the channel $\alpha$ ($\beta$) in the 
two-body c.m. frame;
$F_\pi$ is the pion decay constant; and
the coupling coefficients $C_{{(\alpha)_I}{(\beta)_I}}$ are summarized in Table~\ref{coef}.
As for the cutoff factors $g_{(\alpha)_I}(q')$, we employ the dipole form with the cutoff $\Lambda_{(\alpha)_I}$,
$g_{{(\alpha)_I}}(q')=[\Lambda_{(\alpha)_I}^2/(\Lambda_{(\alpha)_I}^2+q'^2)]^2$.

\begin{table}[thb]
\caption{The coupling coefficients $C_{{(\alpha)_I}{(\beta)_I}}$. 
Note that $C_{{(\alpha)_I}{(\beta)_I}}=C_{{(\beta)_I}{(\alpha)_I}}$.}
\label{coef}       
\begin{ruledtabular}
\begin{tabular}{ccc}
($\alpha,\beta$) & Total Isospin $I$& $C_{{(\alpha)_I}{(\beta)_I}}$ \\
\hline
($\bar K N,\bar K N$)        &$0$ & $6$ \\
($\bar K N,\pi \Sigma$)      &$0$ & $-\sqrt{6}$ \\
($\pi \Sigma ,\pi \Sigma$)   &$0$ & $8$ \\
($\bar K N,\bar K N$)        &$1$ & $2$ \\
($\bar K N,\pi \Sigma$)      &$1$ & $-2$ \\
($\bar K N,\pi \Lambda$)     &$1$ & $-\sqrt{6}$ \\
($\pi \Sigma ,\pi \Sigma$)   &$1$ & $4$ \\
($\pi \Sigma ,\pi \Lambda$)  &$1$ & $0$ \\
($\pi \Lambda ,\pi \Lambda$) &$1$ & $0$ \\
($\pi N ,\pi N$)             &$1/2$ & $4$ \\
($\pi N ,\pi N$)             &$3/2$ & $-2$ 
\end{tabular}
\end{ruledtabular}
\end{table}

It is noted that except for the cutoff factors, both of the above potentials 
[Eqs.~(\ref{eq:e-indep}) and~(\ref{eq:e-dep})]
are derived from the so-called Weinberg-Tomozawa term \cite{Weinberg:1966kf,Tomozawa:1966jm},
which is the leading-order term of the effective chiral Lagrangian,
\begin{equation}
L_{\text{WT}}=\frac{i}{8F_\pi^2}{\rm tr}(\bar{\psi}_B\gamma^\mu[[\phi,\partial _\mu\phi],\psi_B] ),
\label{eq:WT}
\end{equation}
with $\psi_B$ ($\phi$) being the octet baryon (pseudoscalar meson) field.
From this Lagrangian, the $s$-wave potential is given by
\begin{align}
V_{\text{WT}}^{\text{s-wave}} &= 
-\frac{C_{{(\alpha)_I}{(\beta)_I}}}{32\pi^2 F_\pi^2\sqrt{\omega_\alpha(q')
 \omega_\beta(q)}}\sqrt{\frac{(E_\alpha(q')+M_\alpha)(E_\beta(q)+M_\beta)}
 {2E_\alpha(q') 2E_\beta(q)}}\nonumber\\
 &\times
\left[\omega_\alpha(q')+E_\alpha(q')-M_\alpha+\omega_\beta(q)+E_\beta(q)-M_\beta\right],
\label{eq:WT-pot}
\end{align}
where $\omega_\alpha(q')$ [$E_\alpha(q')$] is the meson [baryon] energy of the channel $\alpha$.
We then obtain the E-indep potential~(\ref{eq:e-indep}) from Eq.~(\ref{eq:WT-pot}) 
by assuming $|\vec q~'| \ll m_\alpha, M_\alpha$ and $|\vec q| \ll
m_\beta, M_\beta$.
On the other hand, the E-dep potential~(\ref{eq:e-dep}) 
is given by first replacing $\omega_\alpha(q')+E_\alpha(q')$ and $\omega_\beta(q)+E_\beta(q)$
in the brackets of Eq.~(\ref{eq:WT-pot}) with the on-shell two-body scattering energy $E$, which
is now considered to be an independent variable,
and then assuming $|\vec q~'| \ll m_\alpha, M_\alpha$ and $|\vec q| \ll
m_\beta, M_\beta$.
The replacement with the on-shell two-body scattering energy in deriving the E-dep potential 
corresponds to the so-called ``on-shell factorization''~\cite{Oset:1997it}.

As already seen in Sec.~\ref{sec:three-body},
we take the nonrelativistic kinematics for the numerical calculations.
This is because of a problem inherent in the use of energy-dependent two-body potentials 
for the three-body calculations with the relativistic kinematics.
If the relativistic kinematics are used, the total energy of the two-body subsystem 
can become pure imaginary for large spectator momenta~\cite{Ikeda:2010tk}.
However, such a difficulty does not appear if one uses the nonrelativistic kinematics.

\begin{figure*}
 \includegraphics[width=0.32\textwidth,clip]{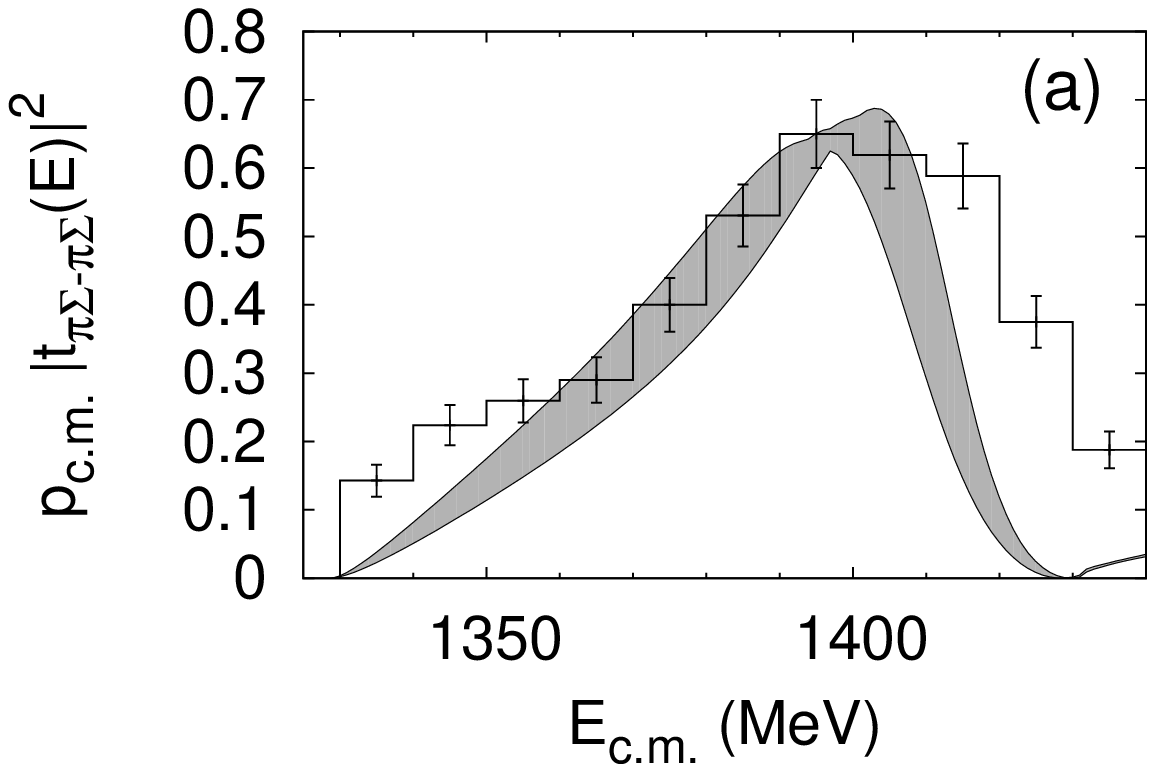}
 \includegraphics[width=0.32\textwidth,clip]{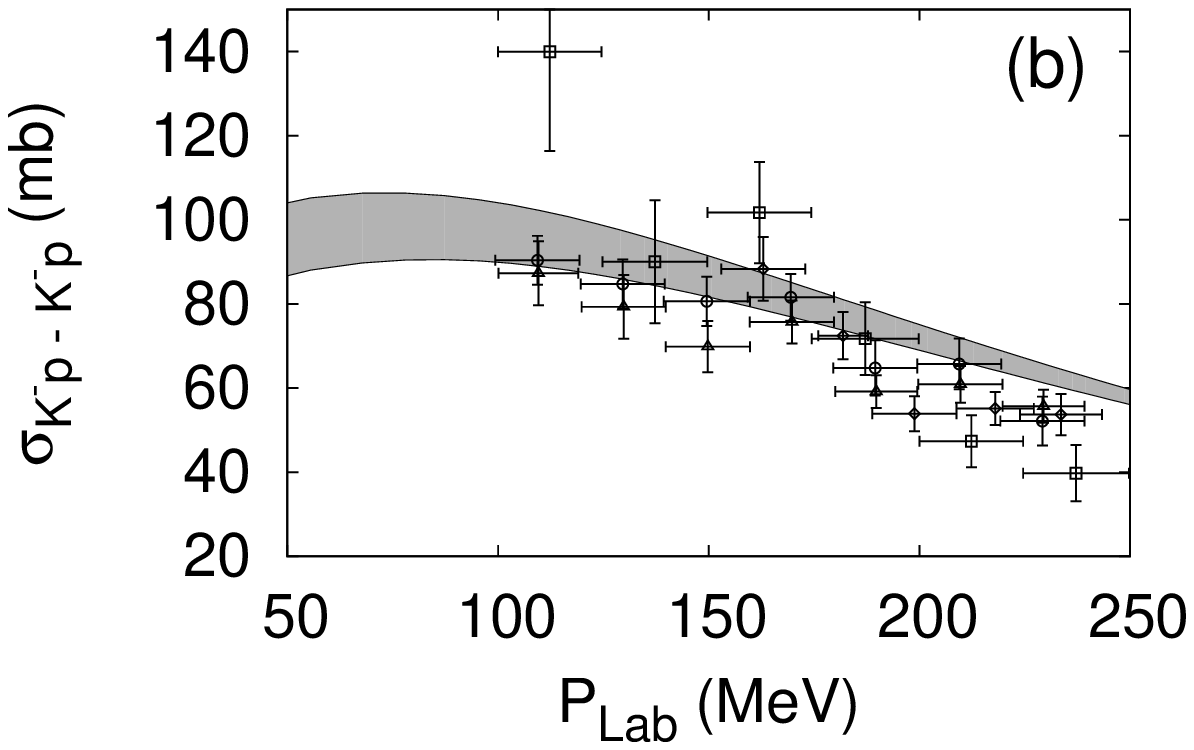}
 \includegraphics[width=0.32\textwidth,clip]{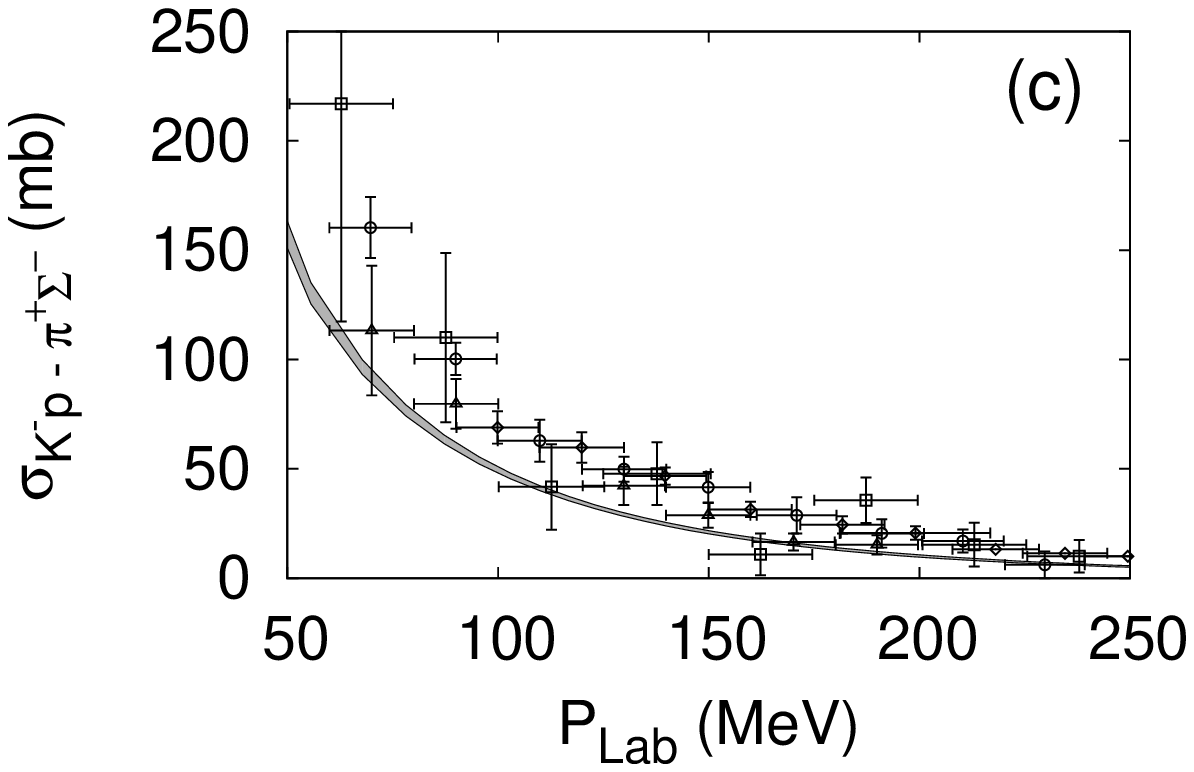}
 \includegraphics[width=0.32\textwidth,clip]{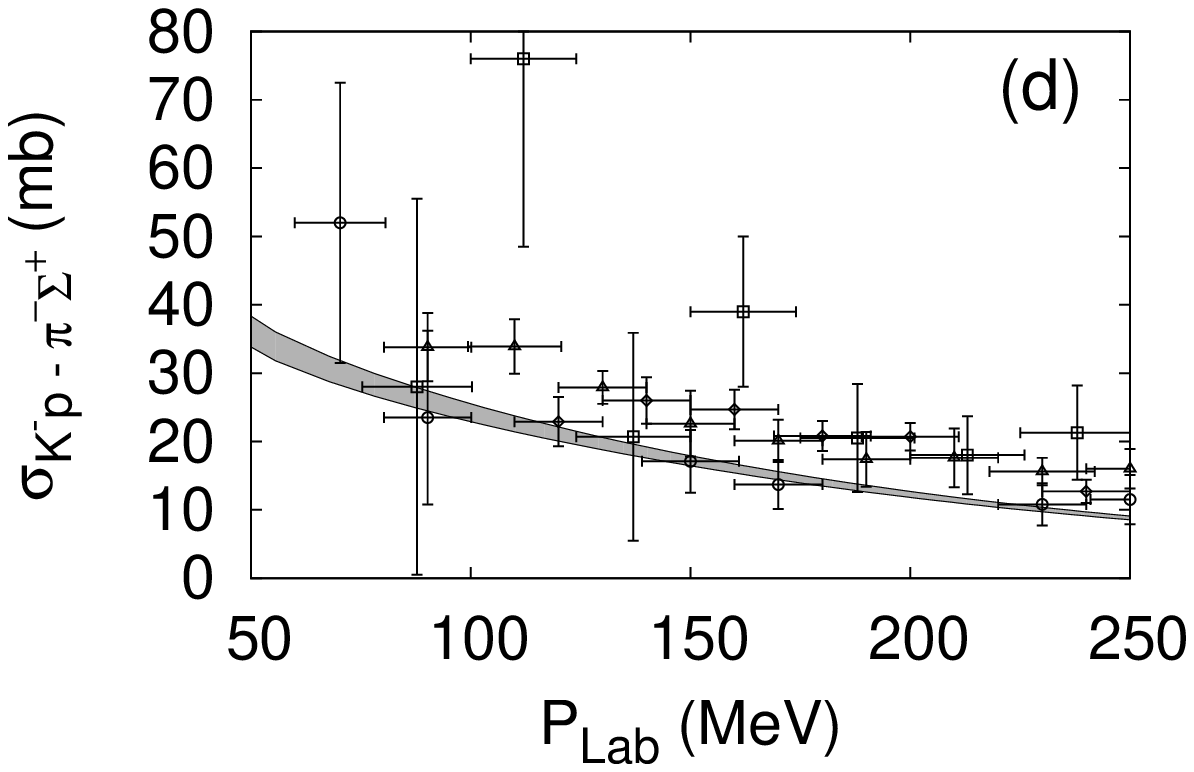}
 \includegraphics[width=0.32\textwidth,clip]{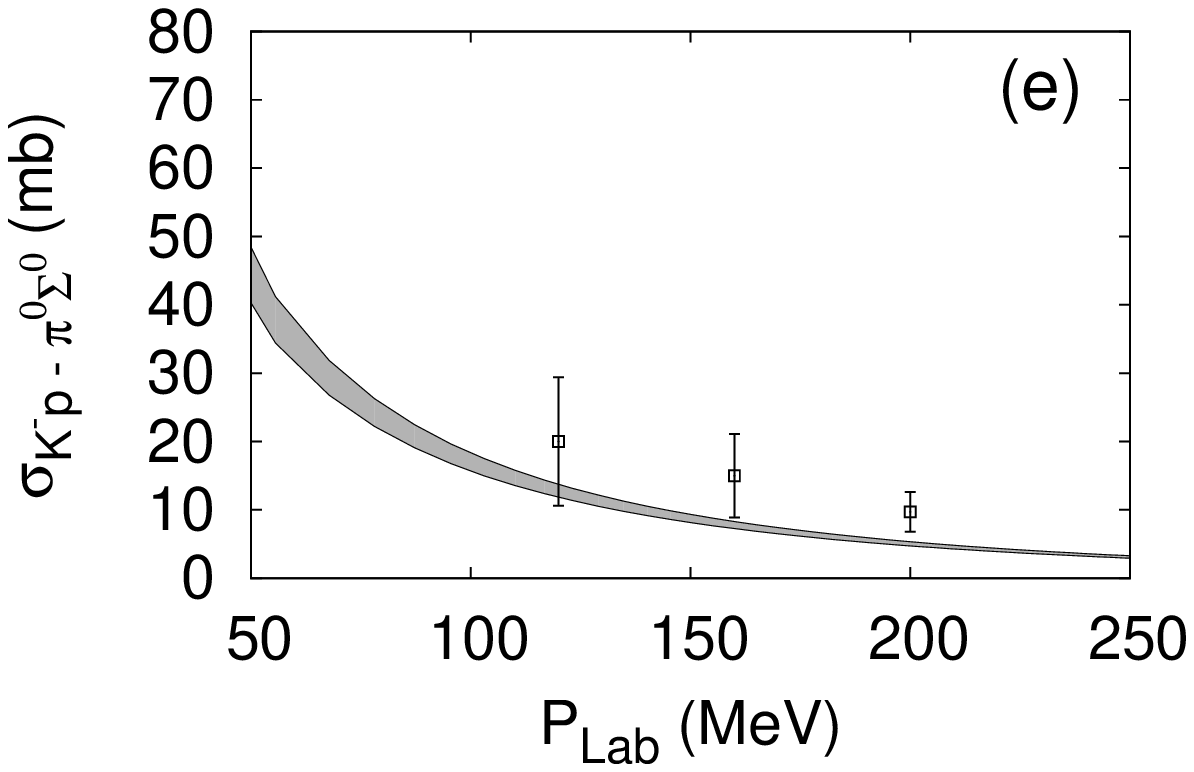}
 \caption{
Results of the fit with the E-indep model.
(a) $I=0$ $\pi\Sigma$ invariant mass distributions of $K^- p \to \pi\pi\pi\Sigma$;
total cross sections of 
(b) $K^-p\rightarrow K^-p$, 
(c) $K^-p\rightarrow \pi^+\Sigma^-$, 
(d) $K^-p\rightarrow \pi^-\Sigma^+$, and 
(e) $K^-p\rightarrow \pi^0\Sigma^0$.
 Data are from Refs.~\cite{Hemingway:1984pz,Humphrey:1962zz,Sakitt:1965kh,Kim:1965zz,Kittel:1966zz,Evans:1983hz}.}
 \label{invariant_indep}
\end{figure*}

\begin{figure*}
 \includegraphics[width=0.32\textwidth,clip]{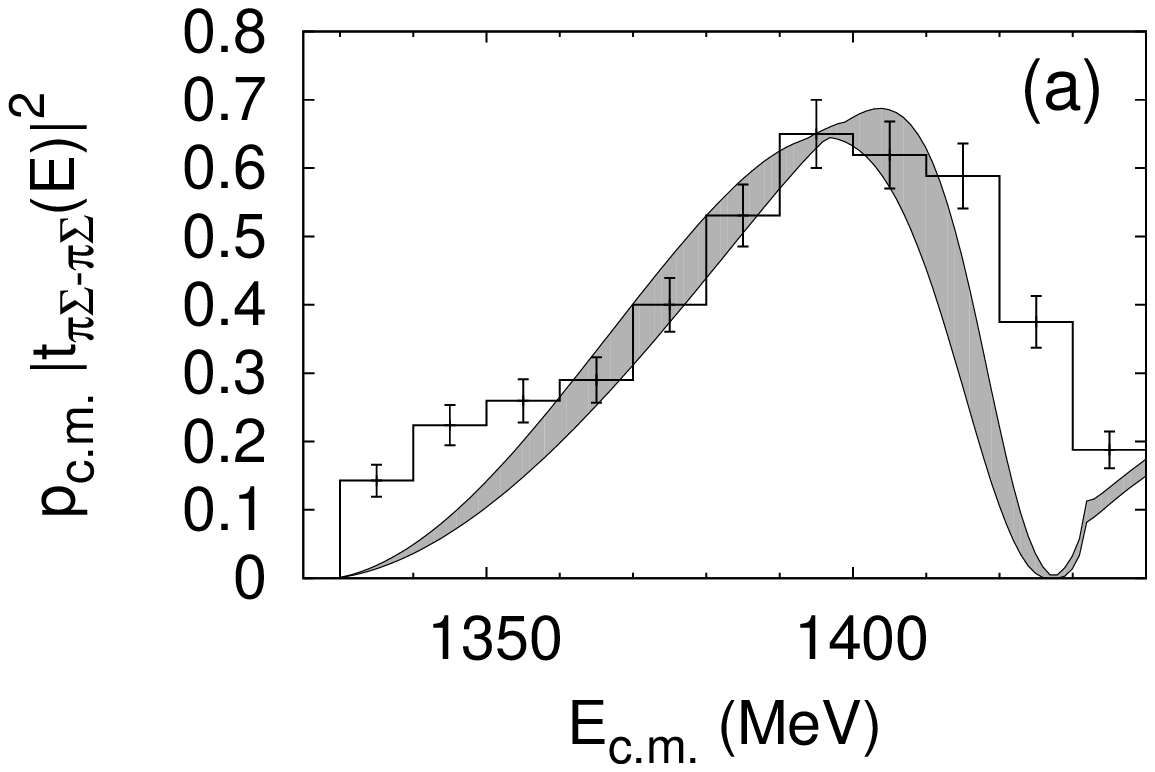}
 \includegraphics[width=0.32\textwidth,clip]{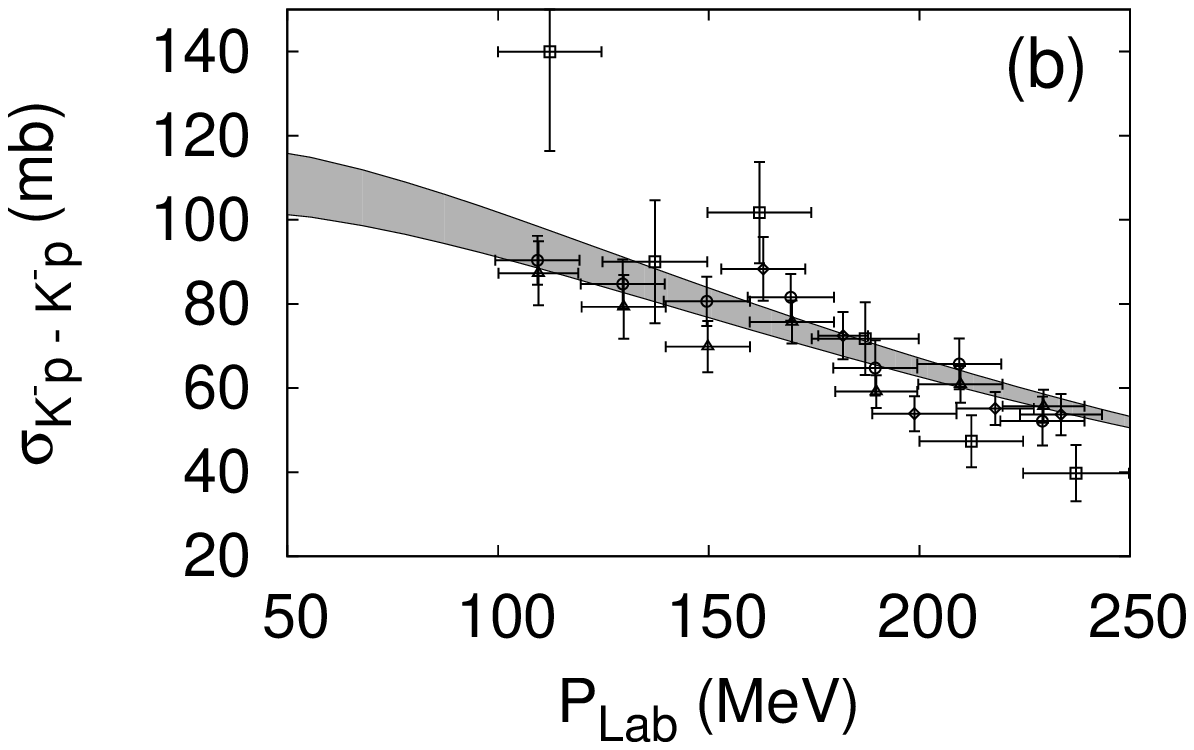}
 \includegraphics[width=0.32\textwidth,clip]{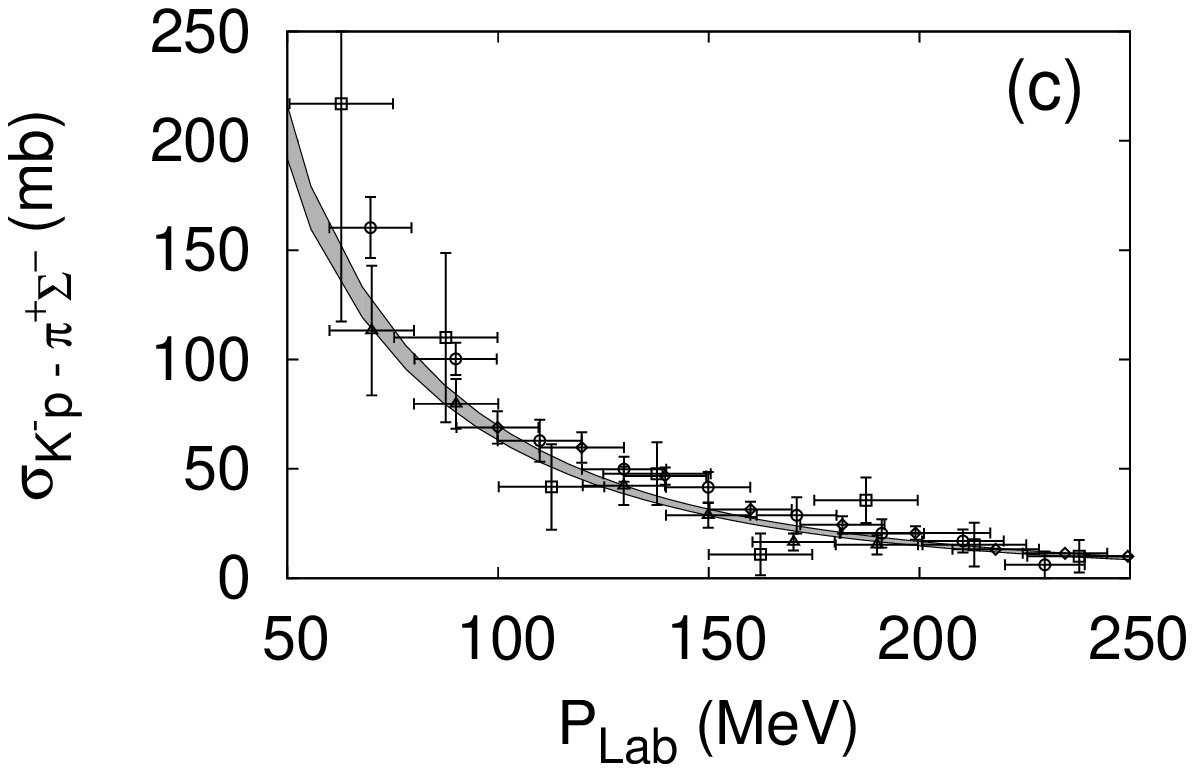}
 \includegraphics[width=0.32\textwidth,clip]{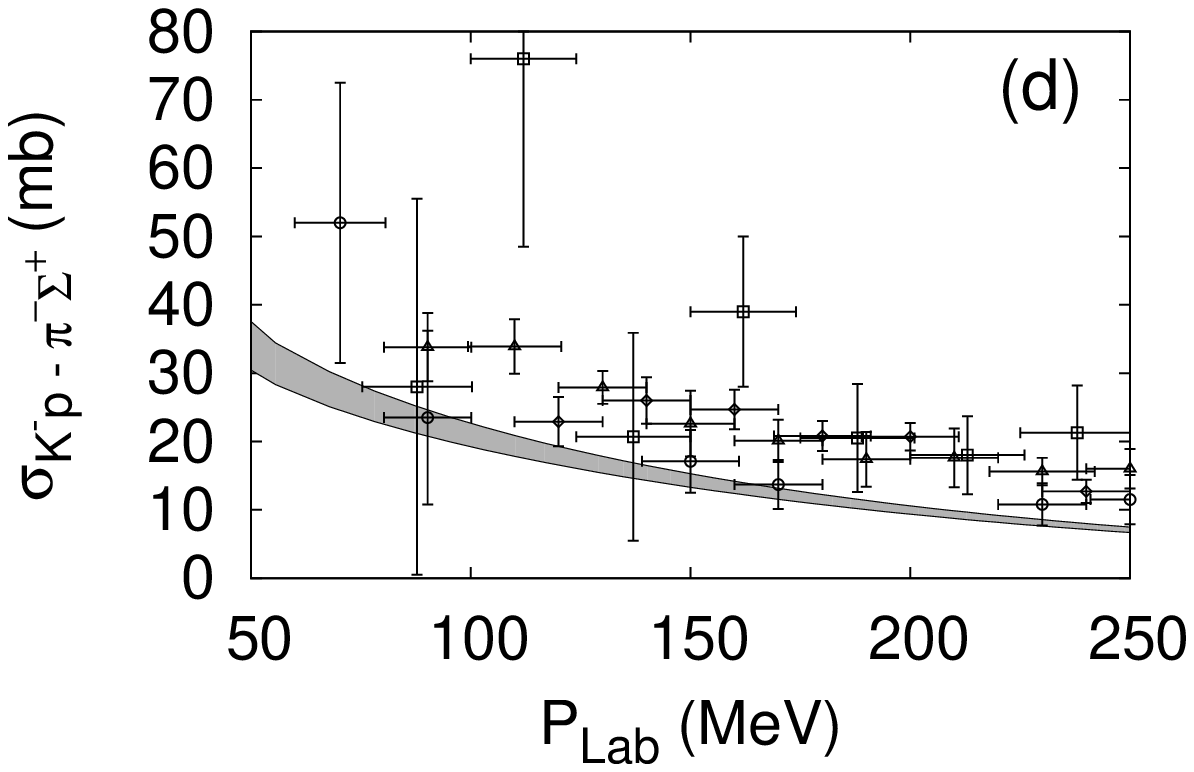}
 \includegraphics[width=0.32\textwidth,clip]{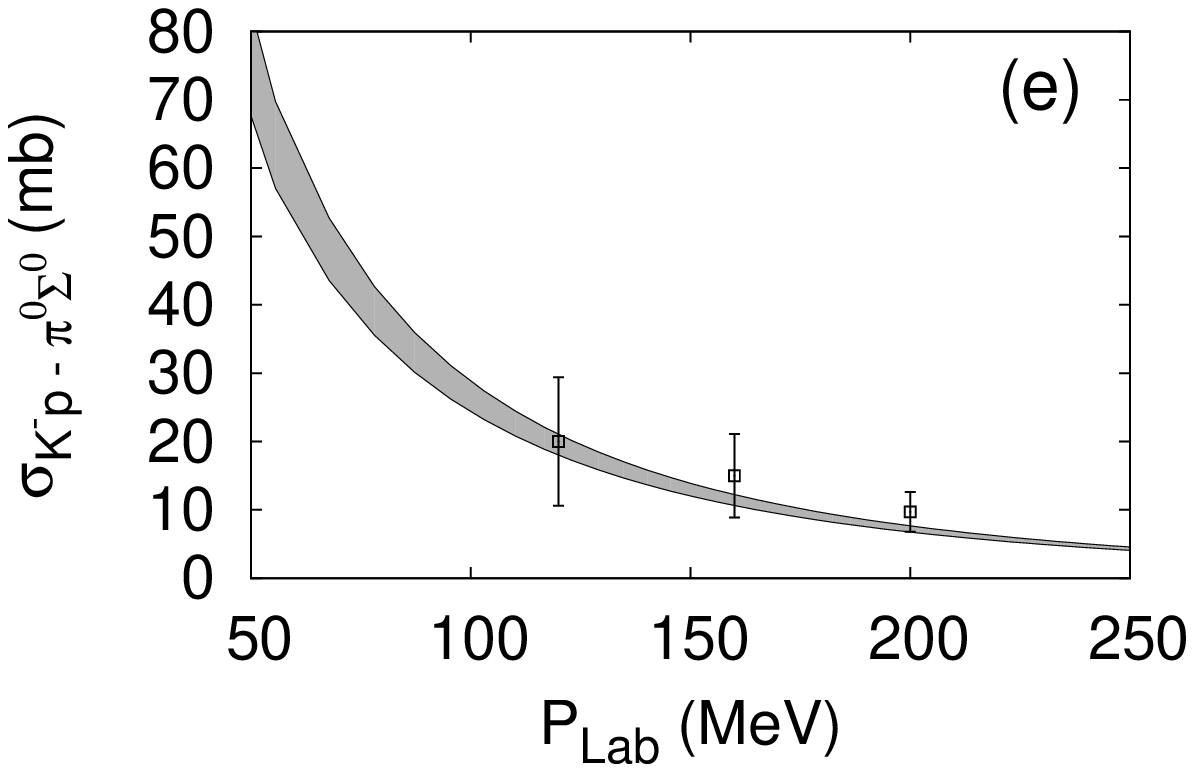}
 \caption{
Results of the fit with the E-dep model.
(a) $I=0$ $\pi\Sigma$ invariant mass distributions of $K^- p \to \pi\pi\pi\Sigma$;
total cross sections of 
(b) $K^-p\rightarrow K^-p$, 
(c) $K^-p\rightarrow \pi^+\Sigma^-$, 
(d) $K^-p\rightarrow \pi^-\Sigma^+$, and 
(e) $K^-p\rightarrow \pi^0\Sigma^0$.
 Data are from Refs.~\cite{Hemingway:1984pz,Humphrey:1962zz,Sakitt:1965kh,Kim:1965zz,Kittel:1966zz,Evans:1983hz}.}
 \label{invariant_dep}
\end{figure*}

Parameters of the two-body potentials are the cutoffs $\Lambda_{(\alpha)_I}$.
We determine the cutoffs 
by fitting the $I=0$ $\pi\Sigma$ invariant mass distributions of 
the $K^- p \to \pi\pi\pi\Sigma$ reaction and the $\bar{K}N$ reaction cross sections.
Results of the fit for the E-indep and E-dep models are presented in 
Figs.~\ref{invariant_indep} and \ref{invariant_dep}, respectively.
There, the results are shown as bands because we have determined
the cutoffs only up to certain ranges within which the computed cross sections are 
consistent with the experimental errors.
The fitted values of the cutoffs are listed in Table~\ref{cutoff}.


\begin{table*}[thb]
\caption{Cutoff parameters of the $\bar{K}N$-$\pi Y$ interaction.}
\label{cutoff}       
\begin{ruledtabular}
\begin{tabular}{lccccc}
 &$\Lambda _{(Y_K)_{I=0}}$ (MeV) &$\Lambda _{(Y_\pi)_{I=0}}$ (MeV) &$\Lambda _{(Y_K)_{I=1}}$ (MeV) &$\Lambda _{(Y_\pi=\pi\Sigma)_{I=1}}$ (MeV)&$\Lambda _{(Y_\pi=\pi\Lambda)_{I=1}}$ (MeV) \\
\hline
E-indep&975-1000 & 675-725 & 920&960&640 \\
E-dep&975-1000 &675-725 & 725&725&725 
\end{tabular}
\end{ruledtabular}
\end{table*}

\begin{figure*}[thb]
\includegraphics[width=0.49\textwidth,clip]{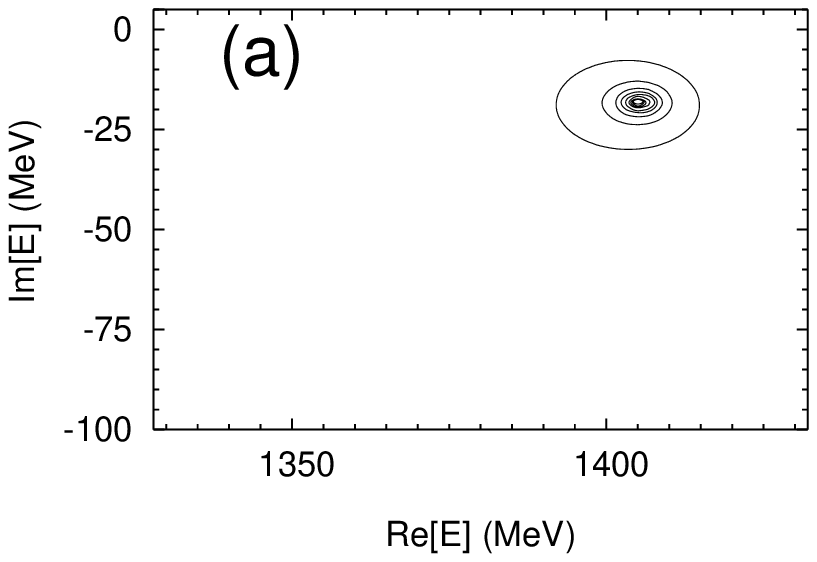}
\includegraphics[width=0.49\textwidth,clip]{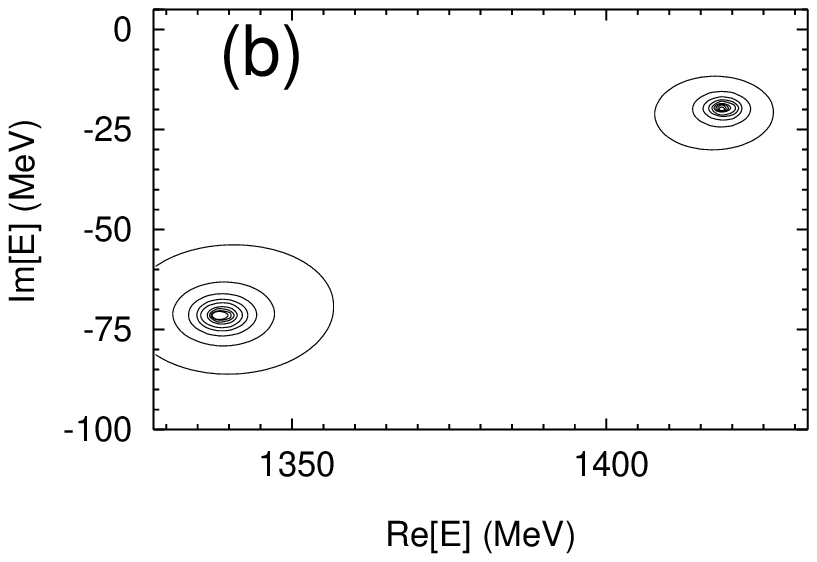}
\caption{
The $S=-1$ and $J^\pi =1/2^-$ $\bar{K}N$ $s$-wave amplitude on 
complex energy plane in
(a) the E-indep model and (b) the E-dep model.
The cutoff parameters are $(\Lambda_{(Y_K)_{I=0}},\Lambda_{(Y_\pi)_{I=0}})=(1000,700)$ MeV.}
\label{pole}
\end{figure*}

In Fig.~\ref{pole}, we present the resonance pole positions of the $\bar K N$ $s$-wave scattering
amplitudes in the complex energy plane between the $\bar{K}N$ and $\pi\Sigma$ threshold energies.
We find that the E-indep model has a single pole
corresponding to $\Lambda(1405)$ in the $\bar{K}N$ physical and 
$\pi\Sigma$ unphysical sheet [Fig.\ref{pole}(a)],
while the E-dep model has two poles in the same sheet [Fig.\ref{pole}(b)]. 
The analytic structure of the amplitudes in the E-dep model is similar to that obtained with
the chiral unitary model~\cite{Jido:2003cb}.

As for the cutoffs with $\alpha = \pi N$, we have determined them by fitting 
the $S_{11}$ and $S_{31}$ $\pi N$ scattering
lengths~\cite{Schroder:1999uq}.
The resulting values are $\Lambda _{(N^*)_{I=1/2}}=\Lambda
_{(N^*)_{I=3/2}}=$400 MeV for both the E-indep and E-dep models.

\subsection{Meson absorption interactions}
\label{sec:abs-int}
To take into account the kaon absorption reaction, we construct the kaon absorption vertex.
In the leading order of the effective chiral Lagrangian, 
there appear interactions associated with the axial-vector couplings.
The interaction Lagrangian is given by
\begin{equation}
L_{\text{abs}}=
-\frac{1}{2F_\pi} \left[
F{\rm tr}(\bar{\psi}_B\gamma^\mu \gamma_5 [\partial _\mu\phi ,\psi_B] )
+ D{\rm tr}(\bar{\psi}_B\gamma^\mu \gamma_5 \{\partial _\mu\phi ,\psi_B \} )
\right]~~,
\label{eq:Yukawa}
\end{equation}
where we employ the empirical values of axial-vector couplings
$F$ and $D$ fixed by the neutron and hyperon decays, i.e.,
$F=0.47$ and $D=0.80$ \cite{Donoghue:1985rk}.
We then find the kaon absorption vertex, $\bar K + N \rightarrow \Lambda $,  as
\begin{equation}
V_{\text{abs}}(\vec p_{\bar{K}}, \vec p_N)=
\frac{i}{\sqrt{6 (2\pi)^3 } F_{\pi}} \sqrt{\frac{1}{2 \omega_{\bar{K}}}} \left( 3F+D \right)
\chi_{s'}^{\dagger}
\left[
 \vec{\sigma} \cdot \vec p_{\bar{K}} -
 \omega_{\bar{K}} \left( \frac{\vec{\sigma} \cdot \vec p_N}{2M_N} + 
\frac{\vec{\sigma} \cdot \left( \vec p_{\bar{K}} + \vec p_N \right) }{2M_{\Lambda}} \right)
\right] \chi_{s}
~~,
\label{eq:V_abs}
\end{equation}
where the $\chi_s (\chi_s')$ and $\vec {\sigma}$ represent the initial (final) nucleon($\Lambda$) spin wave function and the Pauli matrices for the spin.

\subsection{Baryon-baryon interactions}
\label{sec:bb-int}

\begin{figure}[thb]
 \includegraphics[width=0.5\textwidth,clip]{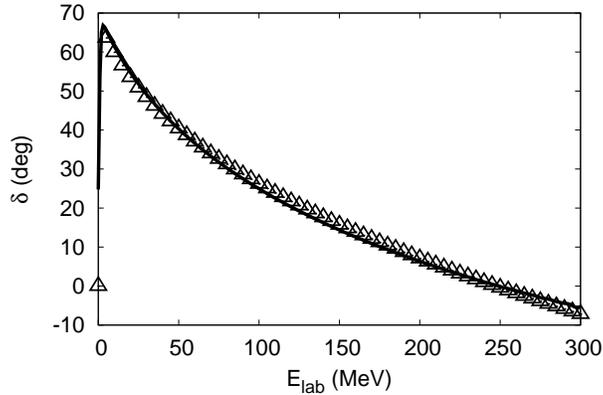}
\caption{Phase shifts of $NN$ scattering for the $^1S_0$
 state. The solid line shows the phase shift
 with our model, and the triangles show the phase shifts
 with the model of Ref.~\cite{Stoks:1994wp}.}
\label{phase_nn}
\end{figure}

As for the s-wave $NN$ interactions, we take the following form~\cite{Ikeda:2007nz}:
\begin{equation}
V_{(d)_{I=1},(d)_{I=1}}({ q}',{ q})=4\pi C_Rg_R({ q}')g_R({ q})-4\pi C_Ag_A({ q}')g_A({ q}).
\end{equation}
Here, $C_R$ ($C_A$) is the coupling strength of the repulsive (attractive) potential.
The form factors $g_{R,A}({ q})$ are defined by 
$g_{R,A}({ q})={\Lambda_{R,A}}^2/({ q}^2+{\Lambda_{R,A}}^2)$, with
$\Lambda_{R,A}$ being the cutoff parameters of the $NN$ interactions.
The coupling strengths $C_{R,A}$ and the cutoff parameters $\Lambda_{R,A}$ are determined by 
fitting the $^1S_0$ phase shifts \cite{Stoks:1994wp} 
(see Fig.~\ref{phase_nn} for the result of the fit).
The resulting values of the parameters are summarized in Table~\ref{cut_nn}.

\begin{table}[thb]
\caption{Parameters of the $NN$ interaction.}
\label{cut_nn}       
\begin{ruledtabular}
\begin{tabular}{cccc}
$\Lambda_R$(MeV) &$\Lambda_A$(MeV) &$C_R$(MeV fm$^3$)&$C_A$(MeV fm$^3$) \\[3pt]
\hline
1215&352&5.05&5.84 
\end{tabular}
\end{ruledtabular}
\end{table}

As for the $s$-wave $YN$ interactions, we follow the form given in 
Ref.~\cite{Torres:1986mr},
\begin{equation}
	V_{(\alpha)_I,(\beta)_I}({ q}',{ q})=-4\pi \frac{C_{(\alpha)_I
	 (\beta)_I}}{2\pi^2}(\mu_\alpha\mu_\beta \Lambda_{(\alpha)_I} \Lambda_{(\beta)_I})^{-1/2}
	g_{(\alpha)_I}({ q}')g_{(\beta)_I}({ q}).
	\label{yn-pot}
\end{equation}
Here, $C_{(\alpha)_I(\beta)_I}$ are the coupling constants 
summarized in Table~\ref{coupl_yn};
$\mu_{\alpha}$ is the reduced mass for the $YN$ system; 
the form factor $g_{(\alpha)_I}({ q})$ is defined by 
$g_{(\alpha)_I}({ q})=\Lambda^2_{(\alpha)_I}/({ q} ^2+\Lambda^2_{(\alpha)_I})$;
and the cutoff parameters $\Lambda_{(\alpha)_I}$ are given by
$\Lambda_{(\Sigma N)_I}=251$ MeV and $\Lambda_{(\Lambda N)_I}=262$ MeV.

\begin{table*}[thb]
\caption{Coupling constants of the $Y N$ interactions.}
\label{coupl_yn}       
\begin{ruledtabular}
\begin{tabular}{cccc}
$C_{(\Sigma N)_{I=1/2}(\Sigma N)_{I=1/2}}$ &$C_{(\Sigma N)_{I=1/2}(\Lambda N)_{I=1/2}}$ &$C_{(\Lambda N)_{I=1/2}(\Lambda N)_{I=1/2}}$ &$C_{(\Sigma N)_{I=3/2}(\Sigma N)_{I=3/2}}$ \\[3pt]
\hline
0.83&0.56&0.49&-0.29
\end{tabular}
\end{ruledtabular}
\end{table*}

\section{RESULTS AND DISCUSSION}
\label{sec:result}

\subsection{Quasi-two-body scatterings}

\begin{figure}[thb]
\includegraphics[width=0.7\textwidth,clip]{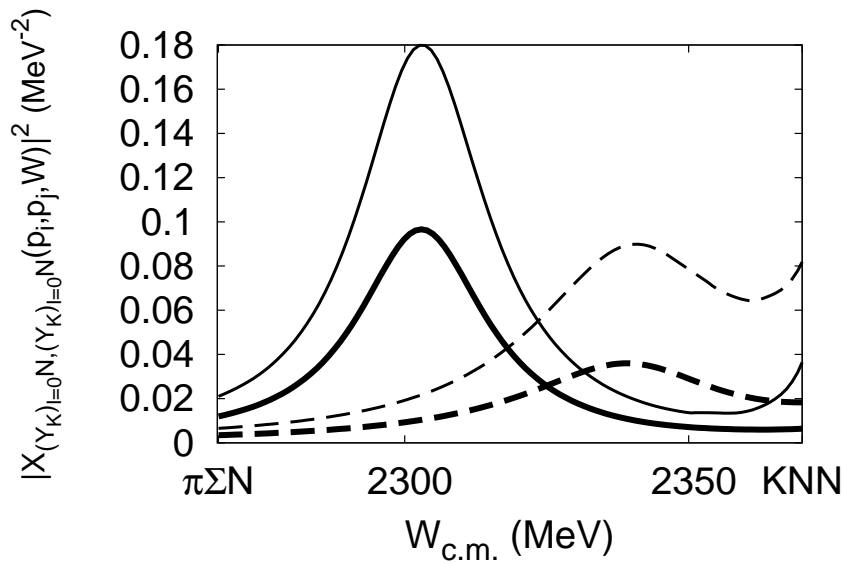}
\caption{$W$ dependence of $|X_{(Y_K)_{I=0}N,(Y_K)_{I=0}N}(p_i,p_j,W)|^2$.
The solid curve is the E-indep model;
the dashed curve is the E-dep model;
the thick curve is $p_i=p_j=150$ MeV;
and the thin curve is $p_i=p_j=100$ MeV.
The cutoff parameters are taken to be 
$(\Lambda_{(Y_K)_{I=0}},\Lambda_{(Y_\pi)_{I=0}}) = (1000, 700)$ MeV.
}
\label{fig:x^2_k}
\end{figure}
\begin{figure*}[thb]
\includegraphics[width=0.45\textwidth,clip]{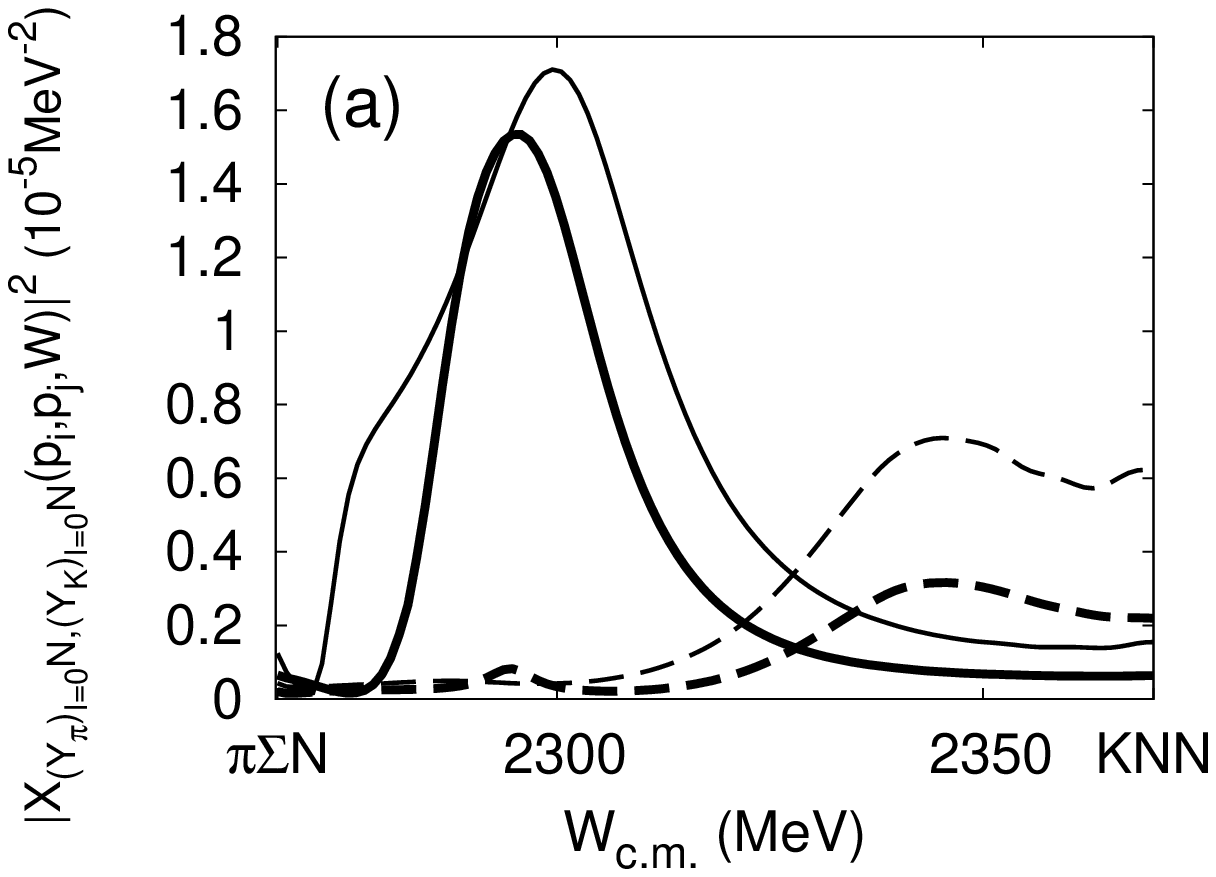}
\includegraphics[width=0.45\textwidth,clip]{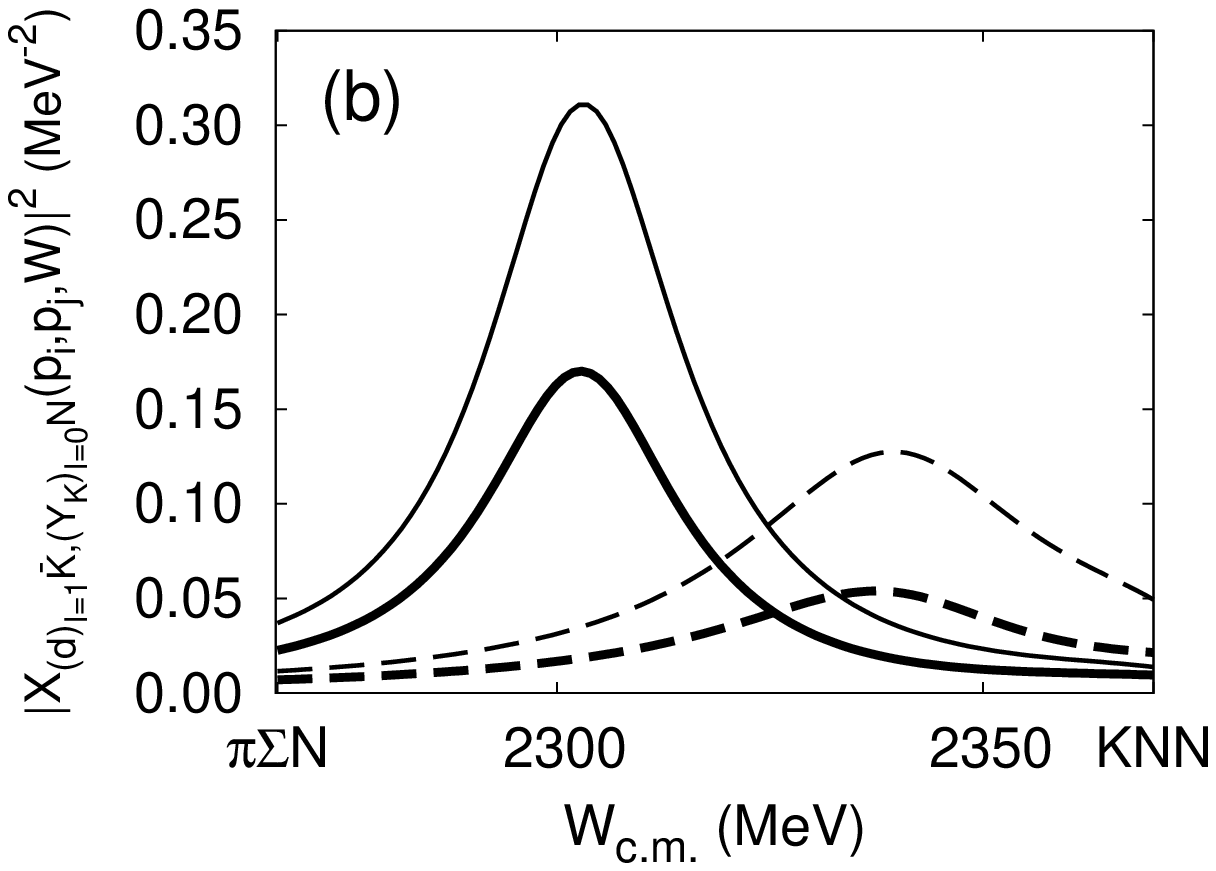}
\includegraphics[width=0.45\textwidth,clip]{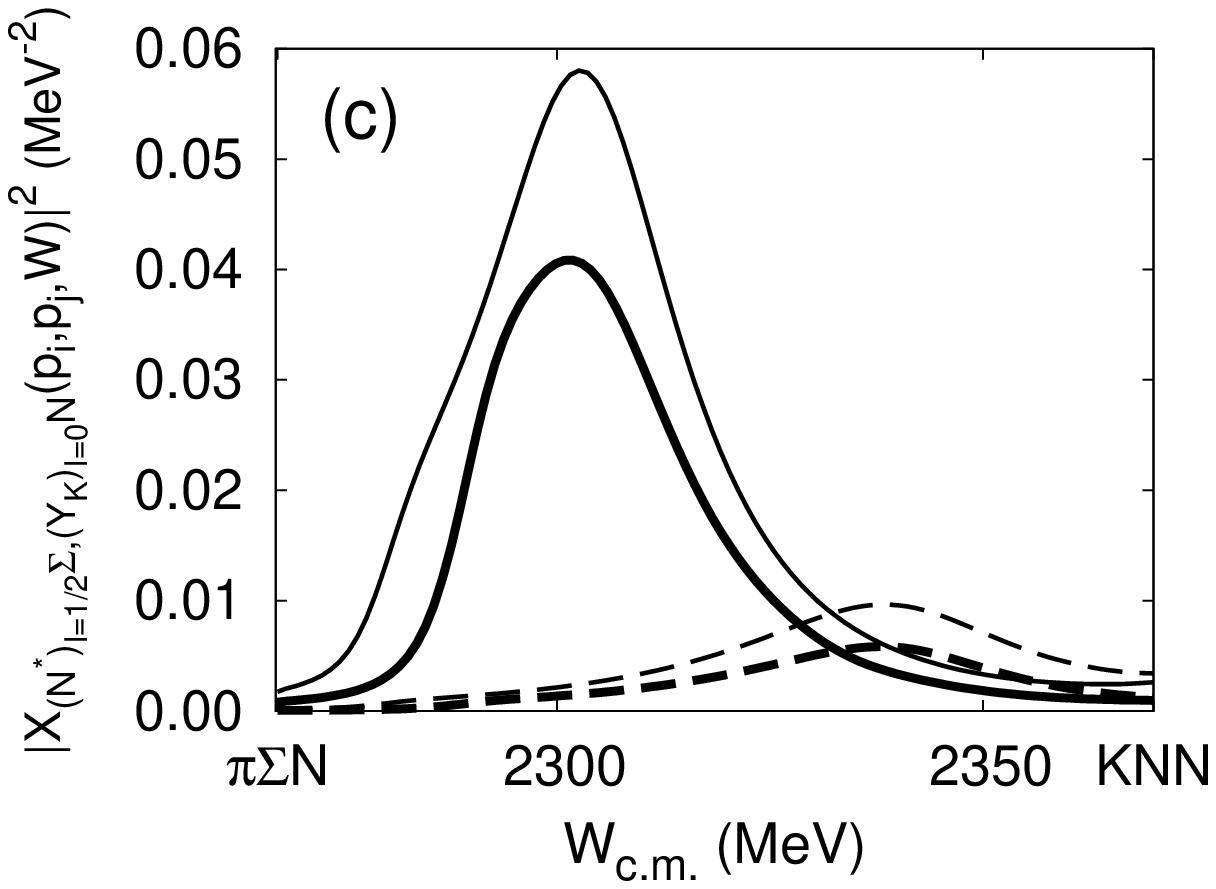}
\includegraphics[width=0.45\textwidth,clip]{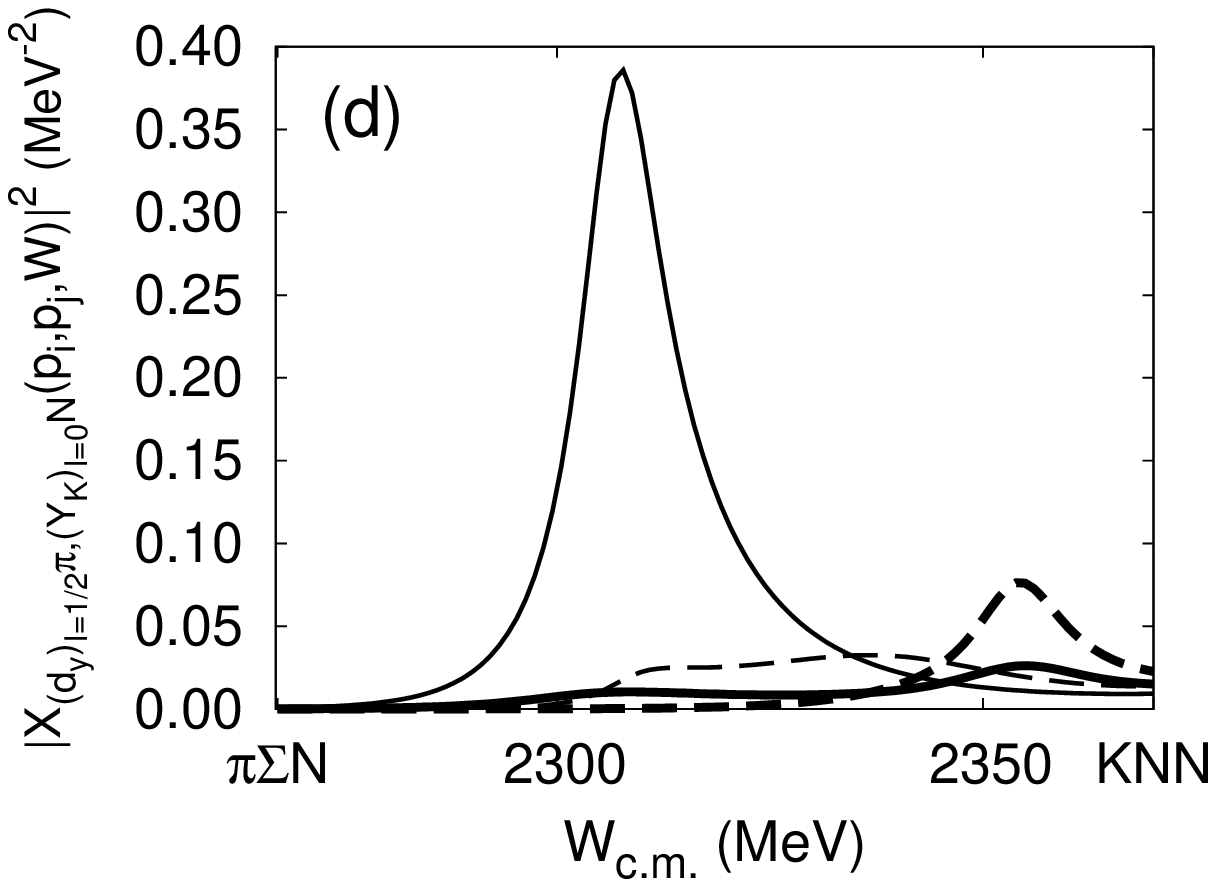}
\caption{$W$ dependence of $|X_{(\alpha)_I i, Y_k(I=0) N}(p_i,p_j,W)|^2$.
(a) $(\alpha)_I=(Y_\pi)_{I=0}$ and $i=N$, 
(b) $(\alpha)_I=(d)_{I=1}(\text{repulsive})$ and $i=\bar K$, 
(c) $(\alpha)_I=(N^*)_{I=1/2} $ and $i= \Sigma$, and
(d) $(\alpha)_I=(d_y)_{I=1/2}$ and $i=\pi$. 
The meaning of each curve and the cutoff parameters 
are taken to be the same as those in Fig.~\ref{fig:x^2_k}.}
\label{fig:x^2_full2}
\end{figure*}

Now we present the partial-wave quasi-two-body amplitudes 
at the real scattering energies $W$, $X_{(\alpha)_Ii,(\beta)_{I'}j}(p_i,p_j,W)$, 
which are obtained by solving the coupled-channel AGS equations 
(\ref{coupled-AGS}) and using the point method explained in the Appendix.
In Fig.~\ref{fig:x^2_k}, we present the absolute square of the amplitudes, 
$|X_{(Y_K)_{I=0}N,(Y_K)_{I=0}N}(p_i,p_j,W)|^2$, whose initial- and final-state isobars are $(Y_K)$ with 
the isospin $I=0$.
Here we plot the results of the E-indep (E-dep) model as solid (dashed) curves.
Also, we plot the amplitudes with two different cases of the off-shell momentum for each model,
with $p_i=p_j=150$ MeV for thick curves and with $p_i=p_j=100$ MeV for thin curves, 
to examine the momentum dependence of the amplitudes.
We find both models have a bump between the $\bar{K}NN$ and $\pi\Sigma N$ threshold energies:
$W\sim 2305$ MeV for the E-indep model and $W\sim 2340$ MeV for the E-dep model,
both of which are close to the resonance pole masses $M_R$ with $-\text{Im}(M_R)\sim 20$ MeV
(see Table~\ref{res_ene}).
Furthermore, the positions of the bumps are independent of the momentum, 
and thus we can conclude that these bumps are actually produced 
by the strange-dibaryon resonances.
On the other hand, in the E-dep model, another strange dibaryon 
with $-\text{Im}(M_R)\sim 100$ MeV barely affects the amplitude on 
the physical real energy axis.
This is consistent with the fact that normally resonances with large widths
cannot produce a sharp peak in the absolute square of the amplitudes or cross sections.
In Fig.~\ref{fig:x^2_full2}, we show the $W$ dependence of the amplitudes with 
different final states.
We observe that the bumps due to the strange-dibaryon resonances appear 
at almost the same $W$ regardless of the final quasi-two-body states, as it should be.
The magnitude of $|X_{(Y_\pi)_{I=0}N,(Y_K)_{I=0}N}(p_i,p_j,W)|^2$ 
[Fig.~\ref{fig:x^2_full2}(a)] is rather small compared with
the other amplitudes shown in Fig.~\ref{fig:x^2_full2}.
This may be understood as follows.
First, as one can notice from the AGS equations (\ref{coupled-AGS}),
the $X_{(Y_\pi),(Y_K)}$  amplitude does not directly couple with
the main $X_{(Y_K),(Y_K)}$ amplitude. 
The $X_{(Y_\pi),(Y_K)}$  amplitude
is generated from the $X_{(N^*),(Y_K)}$ and $X_{(d_y),(Y_K)}$ amplitudes
multiplied by $Z_{(N^*),(Y_K)}\tau_{(N^*),(N^*)}$ and $Z_{(Y_\pi),(d_y)}\tau_{(d_y),(d_y)}$, respectively.
Second, because the $\pi N$ and $YN$ interactions 
are weaker than the $\bar{K}N$-$\pi Y$ and $\bar{K}N$-$\bar{K}N$ interactions,
the $\tau_{(N^*),(N^*)}$ and $\tau_{(d_y),(d_y)}$
propagators are 
 typically
an order of magnitude smaller than
the $\tau_{(Y_K),(Y_K)}$, $\tau_{(Y_K),(Y_\pi)}$, and $\tau_{(Y_\pi),(Y_\pi)}$
that appear in the AGS equations for the other amplitudes.

\begin{figure*}[thb]
\includegraphics[width=0.48\textwidth,clip]{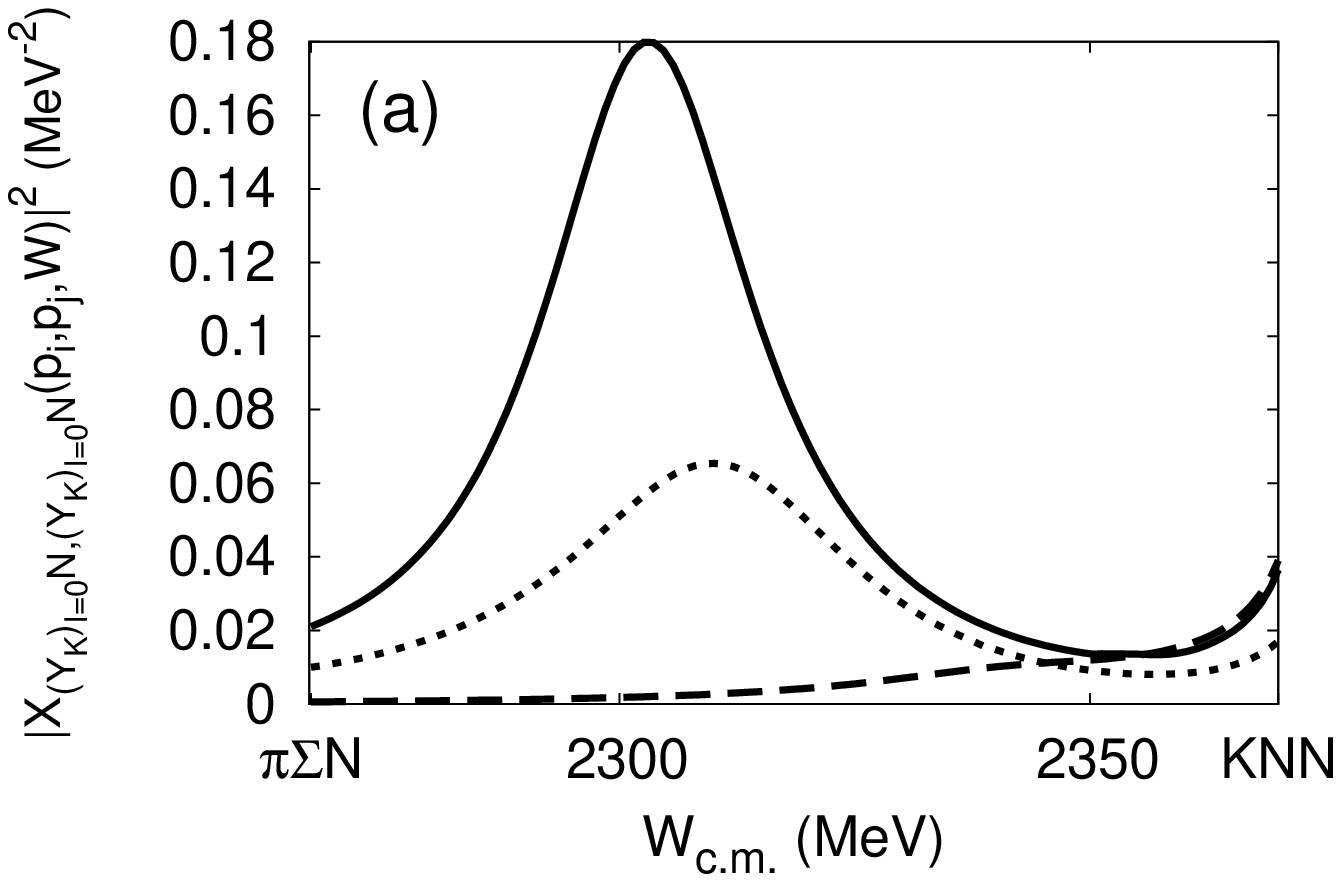}
\includegraphics[width=0.48\textwidth,clip]{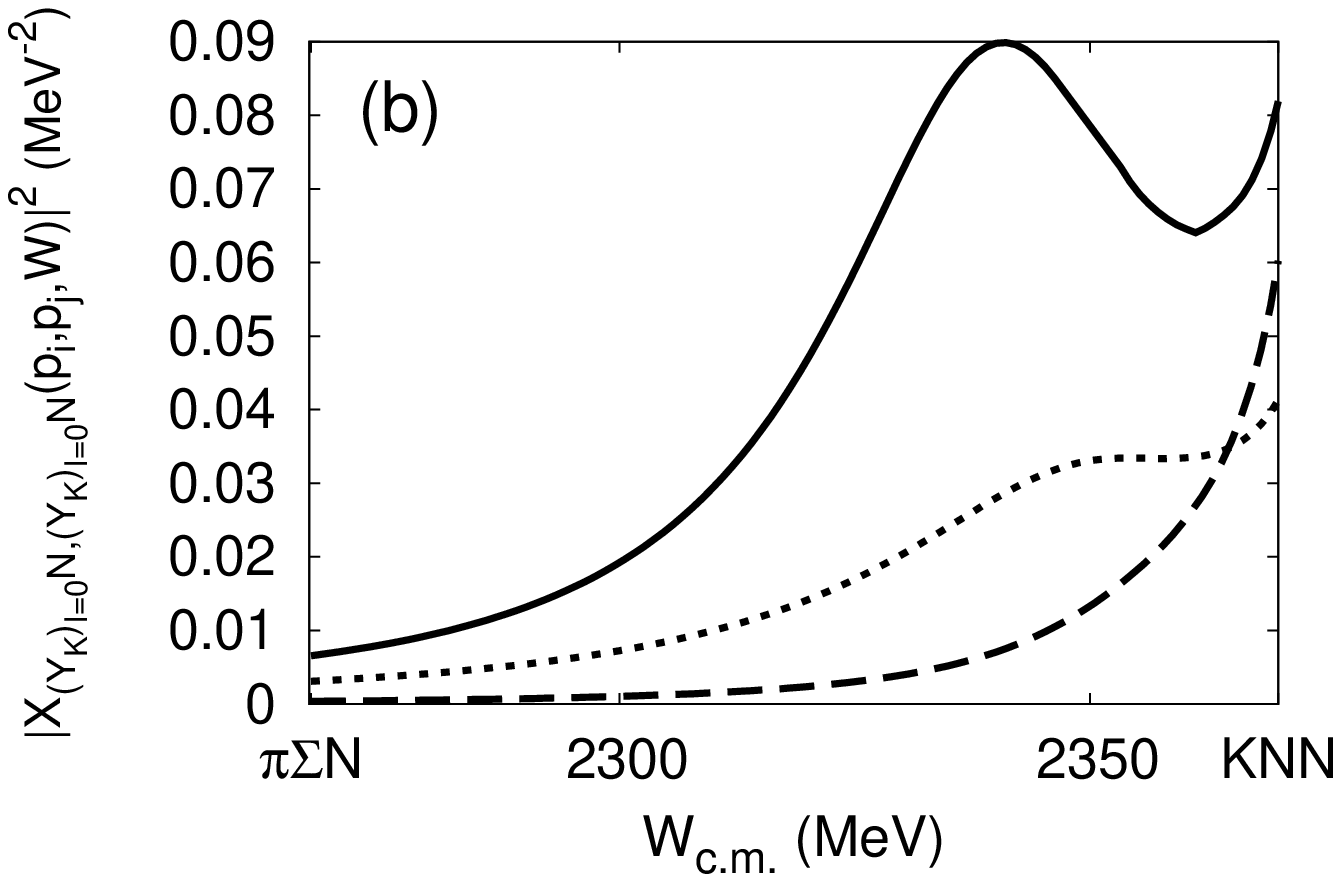}
\caption{
Contributions of one-particle exchange processes to $|X_{(Y_K)_{I=0}N,(Y_K)_{I=0}N}(p_i,p_j,W)|^2$.
The figures are for (a) the E-indep model and (b) the E-dep model. 
The solid curves represent the full results; 
the dashed curves represent the baryon-exchange processes only;
and the dotted curves represent the meson-exchange processes only.
The momentum (cutoff parameters) are fixed as $p_i=p_j=100$~MeV
[$(\Lambda_{(Y_K)_{I=0}},\Lambda_{(Y_\pi)_{I=0}})=(1000,700)$~MeV].
}
\label{fig:x^2_full}
\end{figure*}

Next we present the contributions of each one-particle-exchange mechanism $Z$
to the amplitude $|X_{(Y_K)_{I=0}N,(Y_K)_{I=0}N}(p_i,p_j,W)|^2$ with $p_i=p_j=100$ MeV 
(Fig.~\ref{fig:x^2_full}). 
Here the solid curve in Fig.~\ref{fig:x^2_full}(a) [Fig.~\ref{fig:x^2_full}(b)] 
is same as the thin-solid (thin-dashed) curve in Fig.~\ref{fig:x^2_k}.
If the baryon-exchange (meson-exchange) $Z$ potentials
are switched off in the rescattering processes, then the solid curves 
in Fig.~\ref{fig:x^2_full} are turned into the dashed (dotted) curves.
Contributions of the meson-exchange processes
seem to be crucial for producing the similar bump structure to the full amplitudes, while 
those of the baryon-exchange processes do not.
However, we also observe that rescattering effects including both
the meson- and baryon-exchange processes, which are required by the three-body unitarity,
amplify the magnitude of the scattering amplitudes significantly,
indicating the importance of maintaining the three-body unitarity exactly
in searching for the evidence of the strange-dibaryon resonances.

\subsection{Transition probability for the breakup
$(Y_K)_{I=0}+ N\rightarrow \pi+\Sigma +N$ reaction. 
}

\begin{figure}[thb]
\includegraphics[width=0.7\textwidth,clip]{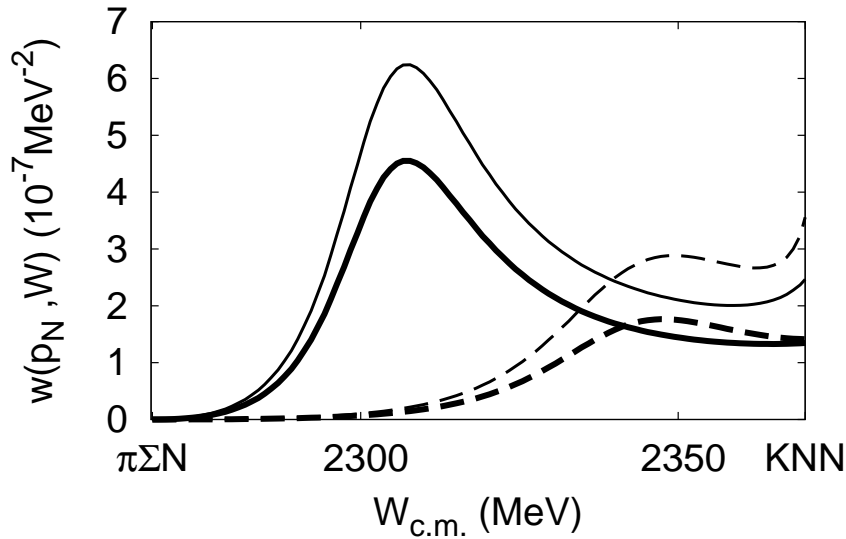}
\caption{
Total transition probability $w(p_N,W)$ for 
$(Y_K)_{I=0}+ N\rightarrow \pi+\Sigma +N$.
The meaning of each curve and the cutoff parameters 
are taken to be same as those in Fig.~\ref{fig:x^2_k}.
}
\label{fig:sig_moment}
\end{figure}

Next, we investigate the energy dependence of the transition probability, $w(p_N,W)$
defined in Eq.~(\ref{eq:cross}), for the $(Y_K)_{I=0}+ N\rightarrow \pi+\Sigma +N$ breakup reaction. 
In Fig.~\ref{fig:sig_moment}, we present $w(p_N,W)$ for $p_N=100$ MeV and $p_N=150$ MeV
using the same values of parameters as used in Fig.~\ref{fig:x^2_k}.
We again find that the position of the bumps in $w(p_N,W)$ are independent of the momentum $p_N$
of the initial $(Y_K) N$ channel, implying that the bumps originate from 
the strange-dibaryon resonances.
The E-indep and E-dep models are found to 
produce quite different energy dependencies on the transition probabilities; 
those differences would be large enough to be detected by experiments.
Because this difference is closely related to
the different nature of $\Lambda(1405)$ between the two models as shown in Fig.~\ref{pole},
the strange-dibaryon production reactions would also provide critical information
on the dynamical origin of $\Lambda(1405)$.

\begin{figure*}[thb]
\includegraphics[width=0.48\textwidth,clip]{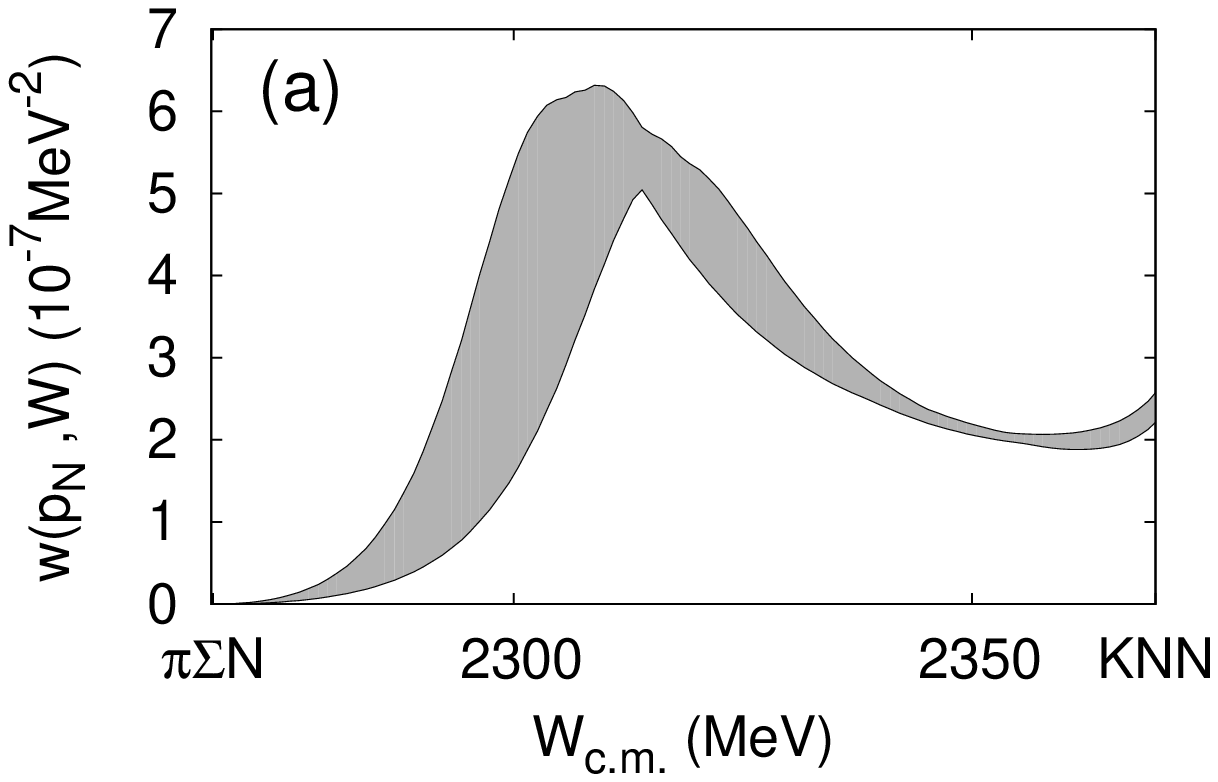}
\includegraphics[width=0.48\textwidth,clip]{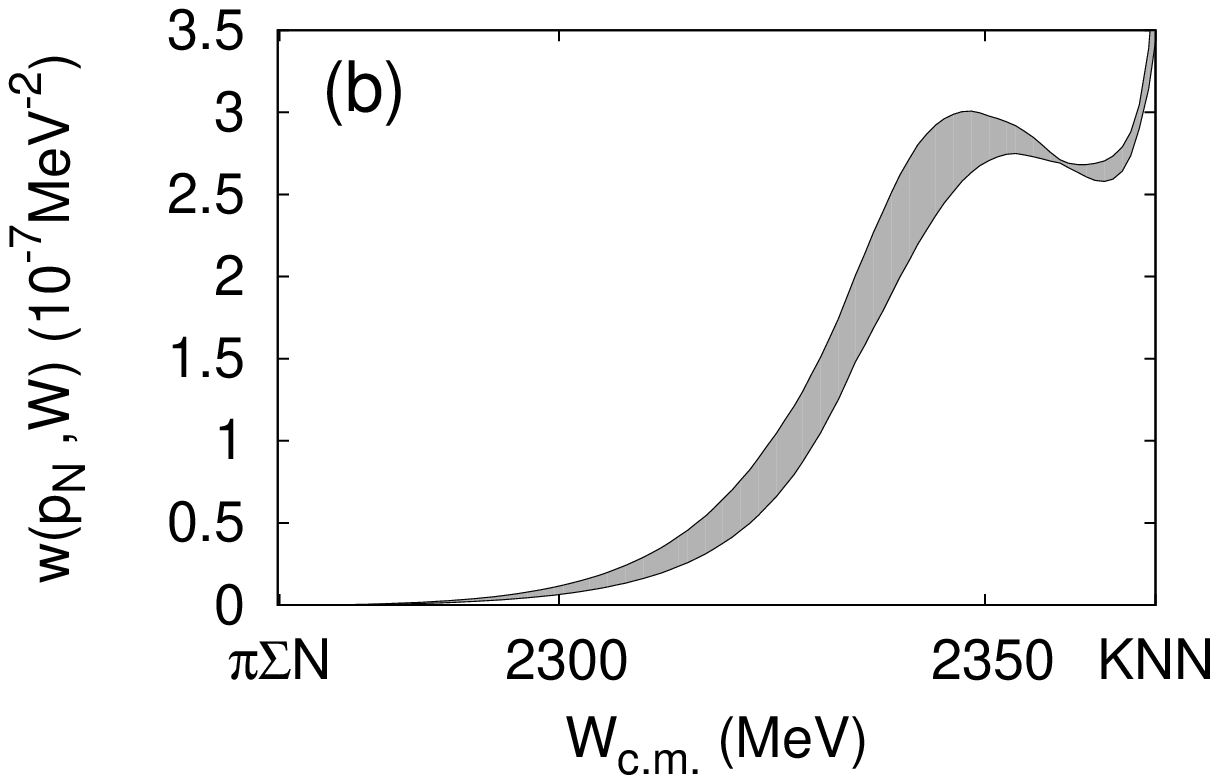}
\caption{
Cutoff dependence on the transition probability for
the $(Y_K)_{I=0}+ N\rightarrow \pi+\Sigma +N$ reaction. 
(a) The E-indep model. (b) The E-dep model.
The bands of transition probability are produced by varying values of
$\Lambda_{(Y_K)_{I=0}}$ and $\Lambda_{(Y_\pi)_{I=0}}$
in the allowed range listed in Table \ref{cutoff}.
The initial nucleon momentum is set to $p_N=100$~MeV.
}
\label{fig:sig}
\end{figure*}

Next we examine the cutoff parameter dependence on the transition probability $w(p_N,W)$
(Fig.~\ref{fig:sig}). 
The bands are given by varying the values of $\Lambda_{(Y_K)_{I=0}}$ and $\Lambda_{(Y_\pi)_{I=0}}$
within the allowed range listed in Table~\ref{cutoff}.
We see that the signal of the strange-dibaryon resonances remains to be observed in
the transition probability within the allowed range 
of $\Lambda_{(Y_K)_{I=0}}$ and $\Lambda_{(Y_\pi)_{I=0}}$.

\begin{figure*}[thb]
\includegraphics[width=0.48\textwidth,clip]{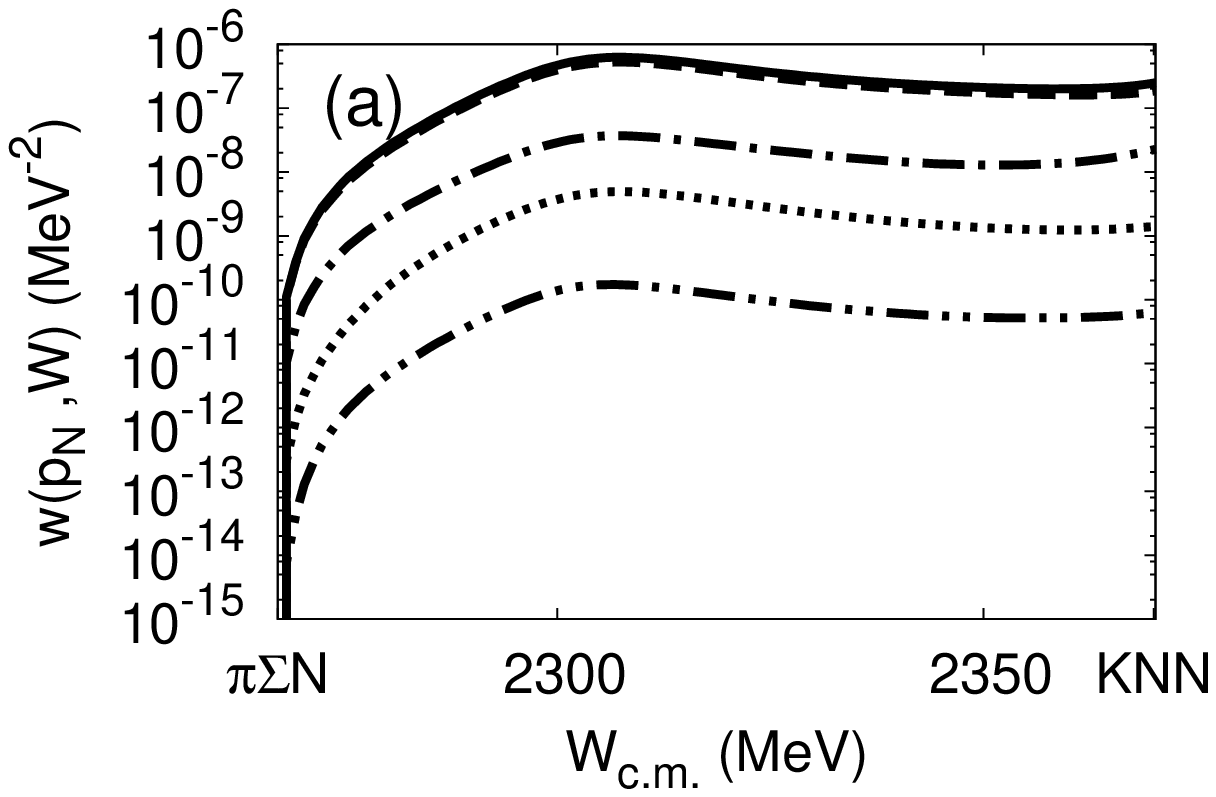}
\includegraphics[width=0.48\textwidth,clip]{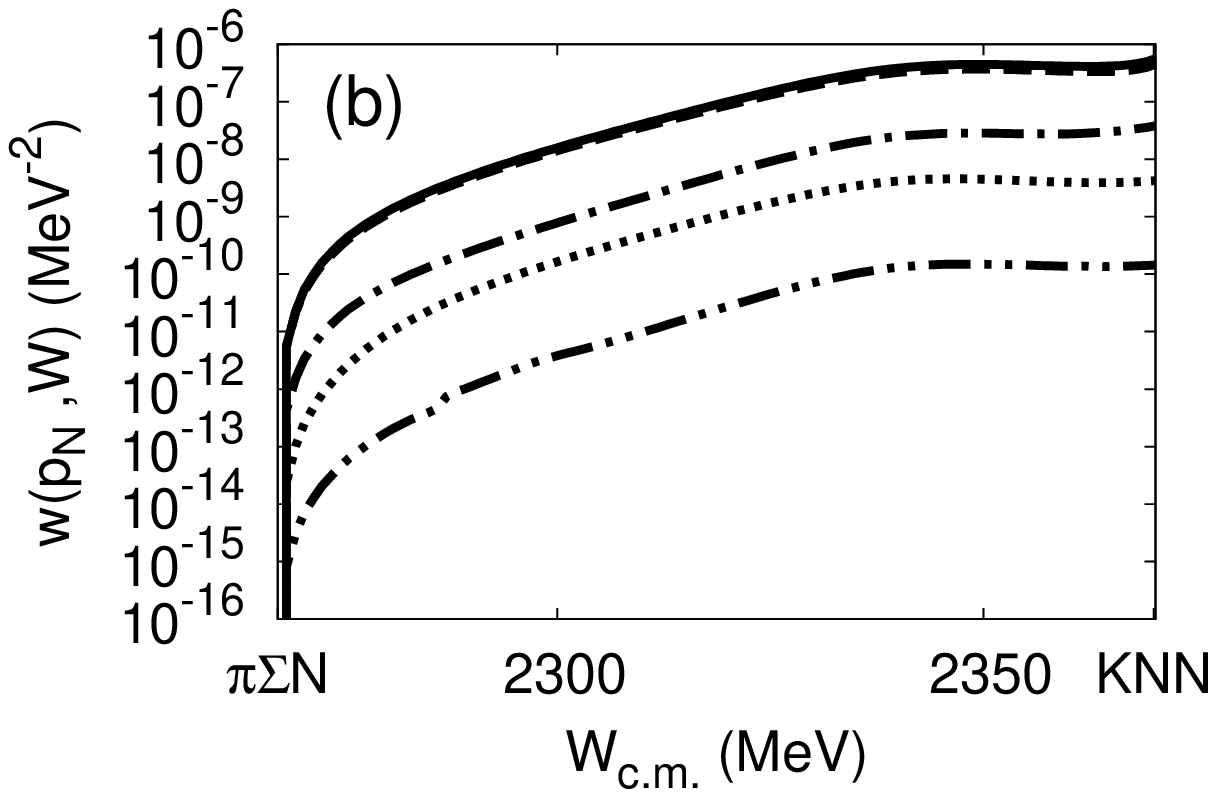}
\caption{
Contribution of each quasi-two-body process to the
$(Y_K)_{I=0}+ N\rightarrow \pi+\Sigma +N$ transition probability.
(a) The E-indep model. (b) The E-dep model.
The solid curve represents the full results;
the dashed curve represents the $X_{(Y_K)_{I} N, (Y_K)_{I=0} N}$ process only;
the dashed-two-dotted curve represents the $X_{(Y_\pi)_I N, (Y_K)_{I=0} N}$ process only;
the dotted curve represents the $X_{(N^\ast)_I\Sigma, (Y_K)_{I=0} N}$ process only;
and the dashed-dotted curve represents the $X_{(d_y)_I\pi, (Y_K)_{I=0} N}$ process only.
The cutoff parameters
$\Lambda_{(Y_K)_{I=0}}$ and $\Lambda_{(Y_\pi)_{I=0}}$ are taken to be 1000 and 700~MeV,
respectively, and the initial nucleon momentum is set to $p_N=100$~MeV.
}
\label{fig:sig_chan}
\end{figure*}

Finally, we examine the contribution of each reaction process to 
the transition probability (Fig.~\ref{fig:sig_chan}).
As can be seen in Eq.~(\ref{eq:t_break}), the reaction processes consist of 
the quasi-two-body processes characterized by the amplitudes 
$X_{(Y_K)_{I} N, (Y_K)_{I=0}N}$, $X_{(Y_\pi)_I N, (Y_K)_{I=0}N}$, $X_{(N^*)_I\Sigma, (Y_K)_{I=0} N}$, and $X_{(d_y)_I \pi, (Y_K)_{I=0} N}$.
We find that the $X_{(Y_K)_{I} N,(Y_K)_{I=0} N}$ process has a dominant
contribution of about 85\%
to the transition probability, while the others have rather small contributions:
about 5\% is from $X_{(d_y)_I\pi, (Y_K)_{I=0} N}$, and
less than 1\% is from $X_{(Y_\pi)_I N, (Y_K)_{I=0} N}$ and $X_{(N^*)_I \Sigma, (Y_K)_{I=0} N}$. 
\subsection{Transition probability for the kaon absorption $(Y_K)_{I=0} + N \rightarrow \Lambda + N$ reaction}

\begin{figure*}[thb]
\includegraphics[width=0.48\textwidth,clip]{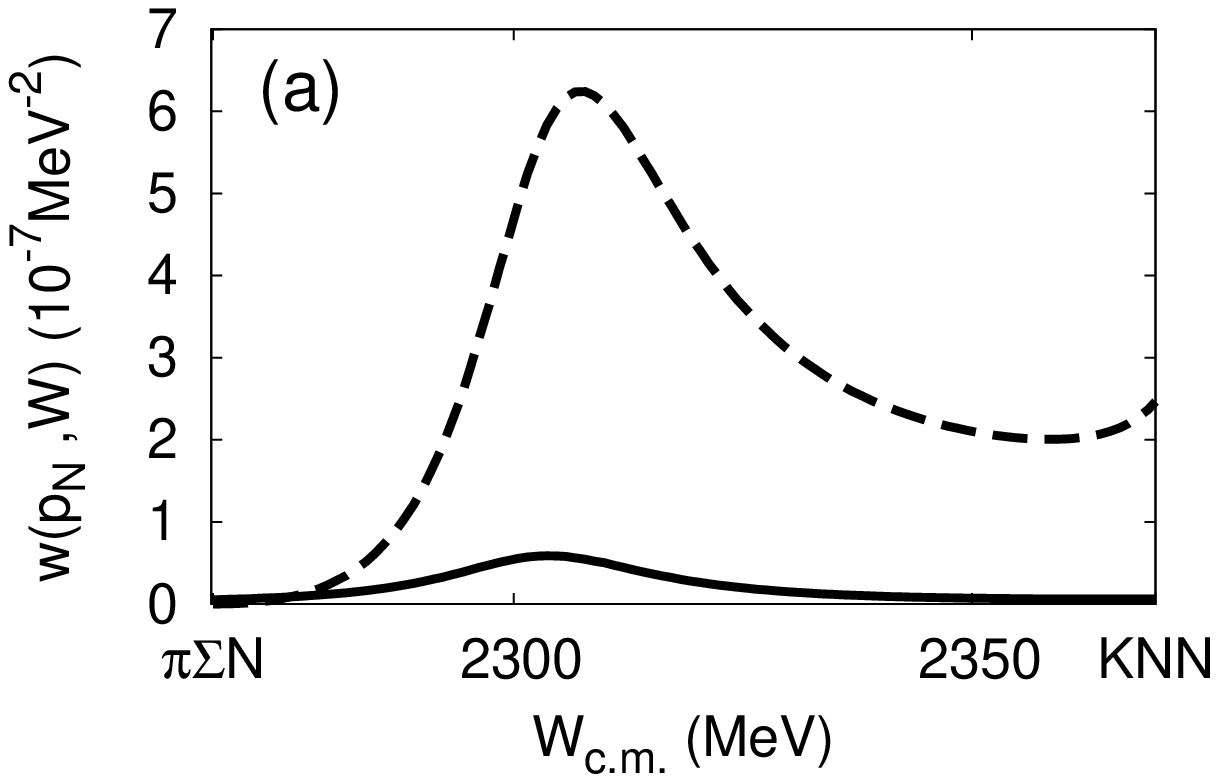}
 \includegraphics[width=0.48\textwidth,clip]{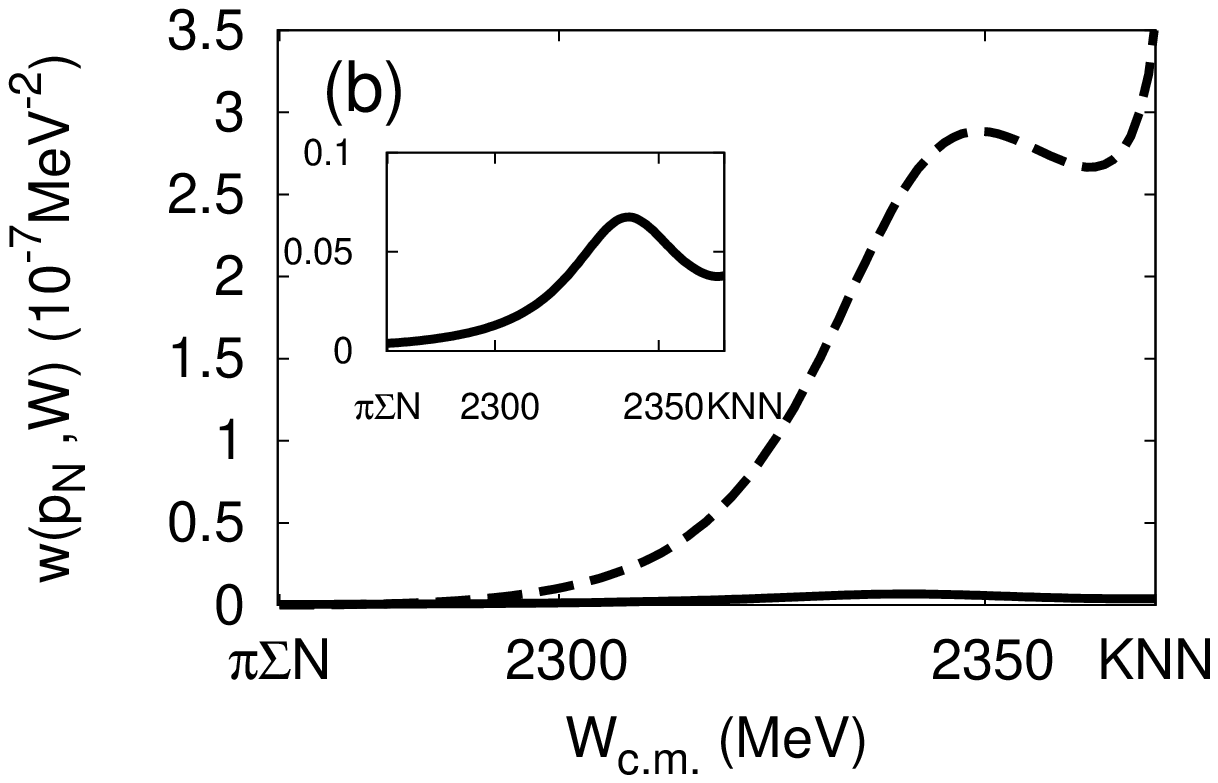}
\caption{
The kaon absorption probability $w_{abs}(p_N,W)$ for
the $(Y_K)_{I=0}+ N\rightarrow \Lambda +N$ reaction (solid lines) for
 (a) the E-indep model and (b) the E-dep model.
Dashed lines represent the transition probability $w(p_N,W)$ for
 $(Y_K)_{I=0}+ N\rightarrow \pi+ \Sigma +N$ reaction. 
The cutoff parameters
$\Lambda_{(Y_K)_{I=0}}$ and $\Lambda_{(Y_\pi)_{I=0}}$ are taken to be 1000 and 700~MeV,
respectively, and the initial nucleon momentum is set to $p_N=100$~MeV.
}
\label{fig:abs}
\end{figure*}
In recent experiments, the $\Lambda p$ channel is used to probe the signal of the 
strange-dibaryon resonances.
We estimate the energy dependence of the transition probability, 
$w_{\rm abs}( p_N,W)$ defined in Eq.~(\ref{eq:w_abs}), for the 
$(Y_K)_{I=0} + N \rightarrow \bar{K}+N+N\rightarrow \Lambda + N$ 
reaction.
As shown in Eq.~(\ref{eq:knn-knn}), in this work we consider
only the $(Y_K)_{I=0}+N\to (Y_K)_{I} + N \to \bar{K}+N+N$ processes
for $(Y_K)_{I=0}+N\to \bar{K}+N+N$.
This is a reasonable simplification because it is found from Fig.~\ref{fig:sig_chan} that
the $X_{(Y_K)_{I}N,(Y_K)_{I=0} N}$ process has the dominant contribution
to the $(Y_K)_{I=0}+ N \rightarrow \pi+ \Sigma+N$ reaction,
and thus we can expect it also for the $(Y_K)_{I=0}+N \to \bar{K}+N+N$ processes.
Then, from Eqs.~(\ref{eq:knn-knn})-(\ref{eq:knn-ln}) and~(\ref{eq:V_abs}),
we can estimate the transition probability $w_{\rm abs}( p_N,W)$.
It is noted that in this work the transition between $\bar{K}+N+N$ and $\Lambda N$ 
is treated perturbatively.
Figure~\ref{fig:abs} shows $w_{abs}(p_N,W)$ for $p_N=100$ MeV,
which is estimated using the same parameter set as that used for calculating 
the quasi-two-body amplitudes $|X_{(Y_K)_{I=0}N,(Y_K)_{I=0}N}(p_i,p_j,W)|^2$ (Fig.~\ref{fig:x^2_k}).
It is found that for both the E-indep and E-dep models 
the bumps due to the strange-dibaryon resonances 
in the $(Y_K)_{I=0}+ N\rightarrow\Lambda+N$ transition probability
become less significant than in the $(Y_K)_{I=0}+ N\rightarrow \pi+ \Sigma +N$ reaction. 
Also, the resonance peak positions move slightly to the downward region from 
the $\bar{K}NN$ threshold energy.

\section{Summary}
\label{sec:summary}
Within the framework of the coupled-channel AGS equations,
we have examined
how the signature of the strange-dibaryon resonances in 
the three-body $\bar{K}NN$-$\pi Y N$ system shows up in 
the scattering amplitudes and transition probabilities on the physical real energy axis.
The logarithmic singularities that appear when solving the AGS equations 
for the real scattering energies
have been successfully handled by making use of the point method.
Two different kinds of models, the E-indep and E-dep models, 
have been considered for the two-body $\bar{K}N$-$\pi \Sigma$ subsystem 
to investigate whether the strange-dibaryon production reactions 
can be used for disentangling the nature of
the two-body $\bar{K}N$-$\pi \Sigma$ system with $\Lambda(1405)$.

We have found that within our model, a clear bump produced by strange-dibaryon resonances
appear in the quasi-two-body scattering amplitudes $X_{(\alpha)_Ii, (Y_K)_{I=0}N}(W)$ and 
the $(Y_K)_{I=0} + N \to \pi+\Sigma+ N$ transition probabilities
in the energy region between the $\bar K NN$ and $\pi\Sigma N$ thresholds,
which strongly suggests that the clear signals of strange-dibaryon resonances 
should be detected by measuring of $\pi\Sigma N$ invariant mass distributions
at the relevant energies.
We have also found that the E-indep and E-dep models
produce quite different energy dependencies on $X_{(\alpha)_I i, (Y_K)_{I=0} N}(W)$ and 
$(Y_K)_{I=0} + N \to \pi+\Sigma+ N$ transition probabilities; those differences 
would be large enough to be detected by experiments.
Within our framework, this difference originates 
from the different nature of $\Lambda(1405)$ between the two models 
as shown in Fig.~\ref{pole}, and thus
the strange-dibaryon production reactions would also be helpful
to reveal the dynamical origin of $\Lambda(1405)$.
We have also studied the spectrum of the $\Lambda N$ final state
by using a simple kaon absorption model.
It was found that the signature of the strange-dibaryon resonances 
in the $\Lambda N$ channel is less significant than that of the three-body final state
due to the stronger contribution of the background amplitudes.

It is for the first time that
the breakup $(Y_K)_{I=0} + N \to \pi+\Sigma+ N$ transition probabilities
are computed within the fully coupled-channel AGS equations.
As a next step, we will further account for initial-state interactions
and develop a technique to make practical calculations of ``actual'' cross sections
of kaon- and photon-induced strange-dibaryon production reactions shown in Fig.~\ref{reaction},
which will be measured at experimental facilities such as J-PARC and SPring-8.
This will be discussed elsewhere.

\begin{acknowledgments}
The simulation has been done on a supercomputer (NEC SX8R) at the Research
 Center for Nuclear Physics, Osaka University.
This work is partly supported by the Yamada Science Foundation.
Y.I. and H.K. acknowledge support from the HPCI Strategic Program (Field 5 ``The
Origin of Matter and the Universe'') of the Ministry of Education, Culture,
Sports, Science and Technology (MEXT) of Japan. 
This work is also supported by JSPS KAKENHI [Grants No. 25800170 (Y.I.), No. 25800149 (H.K.), and No. 24540273 (T.S.)].
\end{acknowledgments}

\begin{appendix}
\appendix			
\section{Brief description of the point method}
\label{sec:3}
\begin{figure}[thb]
\includegraphics[width=0.5\textwidth,clip]{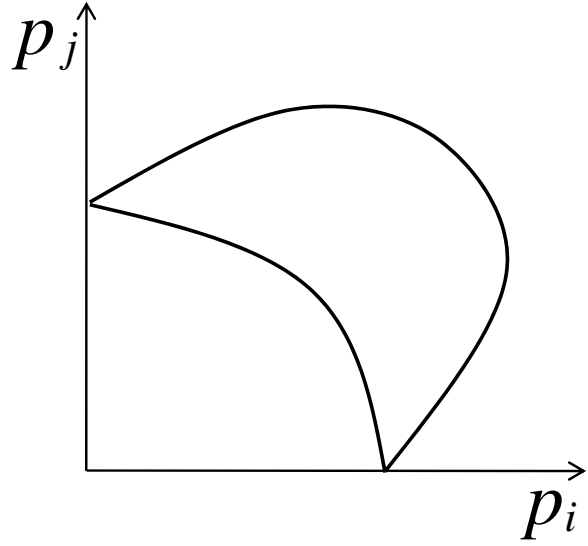}
\caption{
The moon-shaped singularities. The solid curve shows the momentum $(p_i,p_j)$ 
where $Z_{(\alpha)_Ii,(\beta)_{I'}j}(p_i,p_j,W)$ has logarithmic singularity.
}
\label{moon}
\end{figure}

The $s$-wave projection of the particle-exchange potential, $Z_{(\alpha)_Ii,(\beta)_{I'}j}(p_i,p_j,W)$ 
[Eq.~(\ref{eq:z-diagram-s})], contains the following logarithm:
\begin{align}
 \ln\left[\frac{W-M-\frac{p_i^2}{2m_i}-\frac{p_j^2}{2m_j}
 -\frac{p_i^2+p_j^2-2p_ip_j}{2m_k}}{W-M-\frac{p_i^2}{2m_i}
 -\frac{p_j^2}{2m_j}-\frac{p_i^2+p_j^2+2p_ip_j}{2m_k}}\right]~~.
\end{align}
For real $W$ with $W > M$, this logarithm becomes singular at momentum $(p_i,p_j)$ satisfying
\begin{align}
 W-M-\frac{p_i^2}{2m_i}-\frac{p_j^2}{2m_j}-\frac{p_i^2+p_j^2\pm 2p_ip_j}{2m_k}=0~~.
\end{align}
The singularities appear as a ``moon-shape'' 
in the $p_i$-$p_j$ plane as illustrated in Fig.~\ref{moon}.

As a practical technique to handle the moon-shaped singularities 
in solving the scattering equations~(\ref{coupled-AGS}),
we have employed the point method,
which is proposed by Schlessinger~\cite{Schlessinger:1968} and developed by 
Kamada~\textit{et al.}~\cite{Kamada:2003xy}.
We briefly explain the method in the following.

The point method is an extrapolation technique of functions.
With this technique, one can evaluate the value of a function $X(W)$ of real $W$ from
$X(W+i\epsilon_i)$, where $\epsilon_i~(\epsilon=1,2,...)$ is a series of 
positive finites that converges to zero, using the following formulas:
\begin{eqnarray}
X(W)
= \lim_{\epsilon\to 0}
\frac{X(W+i\epsilon_1)}{1+\frac{a_1(\epsilon-\epsilon_1)}{1+\cdots}}
= \lim_{\epsilon\to 0}
\frac{X(W+i\epsilon_1)}{1+}\frac{a_1(\epsilon-\epsilon_1)}{1+}\frac{a_2(\epsilon-\epsilon_2)}{1+}
\cdots~~,\label{eq:cont_frac}
\end{eqnarray}
with
\begin{eqnarray}
a_l=\frac{1}{\epsilon _l-\epsilon _{l+1}}\bigg( 1+
\frac{a_{l-1}(\epsilon _{l+l}-\epsilon _{l-1})}{1+}\frac{a_{l-2}(\epsilon _{l+1}-\epsilon _{l-1})}{1+}
\cdots
\frac{a_1(\epsilon _{l+1}-\epsilon _1)}{1-[X(W+i\epsilon _1) /X(W+i\epsilon _{l+1})]}\bigg)~~.
\label{eq:cont_coeff}
\end{eqnarray}

To illustrate how we get scattering amplitudes $X_{(\alpha)_Ii,(\beta)_{I'}j}(p_i,p_j,W)$ 
for real $W$, we apply the formulas above to the Amado model~\cite{1963PhRv..132..485A},
a simple model for three-boson scatterings. 
The AGS equations for the $s$-wave scattering of a boson $b$ and a two-$b$ bound-state $d$,
$bd\rightarrow bd$, are given by 
\begin{equation}
 X(p',p_0,W)=2Z(p',p,W)+2\int p^2dpZ(p',p,W)\tau(p,W)X(p,p_0,W)\label{amado_x}.
\end{equation}
In the $bd$ CM system, the driving term $Z(p',p,W)$ and the two-body
propagator $\tau(p,W)$ are expressed as
\begin{align}
 Z(p',p,W)&=\frac{1}{2}\int^1_{-1}dx\frac{g_0}{(|\vec{p}~'+\frac{1}{2}\vec{p}|^2+\beta^2)}
 \frac{g_0}{(|\vec{p}+\frac{1}{2}\vec{p}~'|^2+\beta^2)}\nonumber\\
 &\times\frac{1}{W-\frac{p^2}{2m}-\frac{p'^2}{2m}-\frac{(\vec{p}+\vec{p}~')^2}{2m}+i\epsilon},
 \label{amado_z}
\end{align}
\begin{align}
 \tau^{-1}(p,W)&=\left[E_2(p,W)+B+i\epsilon\right]\nonumber\\
 &=\left[1-(E_2(p,W)+B+i\epsilon)\int
 k^2dk\frac{g^2(k)}{\left(B+\frac{k^2}{m}\right)^2
\left(E_2(p,W)-\frac{k^2}{m}+i\epsilon\right)}\right].
\label{amado_tau}
\end{align}
Here, $g(q)=g_0/(q^2+\beta^2)$ is the form factor for $d\rightarrow bb$,
which is normalized as $\int k^2dkg^2(k)/(B+\frac{k^2}{m})^2=1$;
$B$ is the binding energy of $d$; and $E_2(p,W)=W-3p^2/(4m)$ is
the two-body scattering energy.
We solve these AGS equations by setting $\hbar=2m=1$,
$B=1.5$, $\beta=5$, and $W=1$.

If one tries to solve Eq.~(\ref{amado_x}) for a real $W$, the momentum integral path
crosses the singularities of the $Z$ potential and thus the resulting amplitude 
$X(p',p_0,W)$ does not converge.
On the other hand, one can have convergent solutions of Eq.~(\ref{amado_x}) without any problems
for complex energies $W + i\epsilon_l$ with positive finites $\epsilon_l$. 
Therefore, we first compute the amplitude $X$ for several complex energies and then make 
an extrapolation to $X(W)$ using Eqs.~(\ref{eq:cont_frac}) and~(\ref{eq:cont_coeff}).
For practical computations, we use five $\epsilon_l$'s:
\begin{equation}
\epsilon_l=0.05\times l \qquad (l=1,2,\dots,5).
\end{equation}

\begin{figure}[thb]
\includegraphics[width=0.75\textwidth,clip]{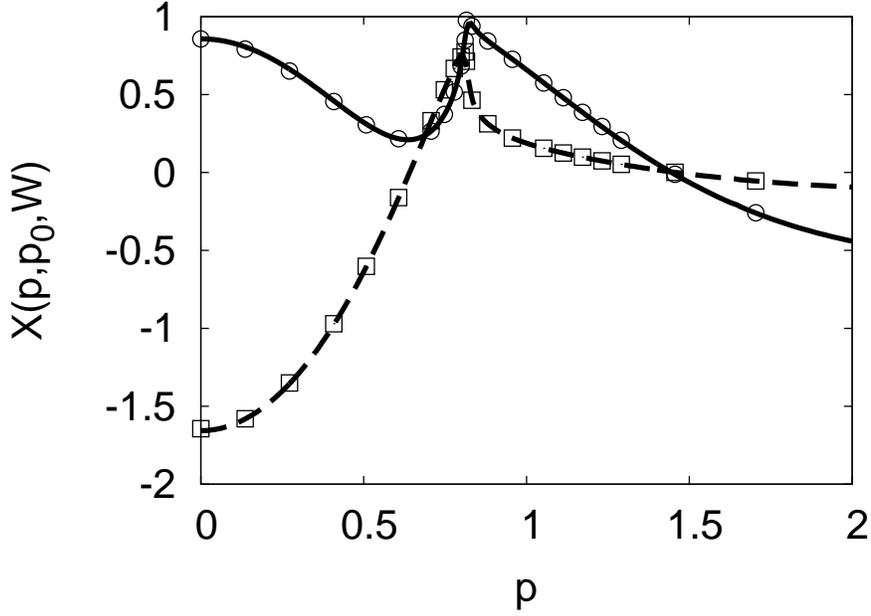}  
\caption{
The amplitude $X(p,p_0,W)$ of the Amado model at $W=1$ and the on-shell momentum 
$p_0=\sqrt{4m(W+B)/3}$. 
Solid and dashed curves are the real and imaginary parts of 
the amplitude $X(p,p_0,W)$, respectively, extrapolated by 
the point method~\cite{Schlessinger:1968,Kamada:2003xy}.
Circles and squares are the real and imaginary parts of the amplitude $X(p,p_0,W)$ by the spline
interpolation method~\cite{Matsuyama:2006rp}.
}
\label{fig:amado}
\end{figure}

In Fig.~\ref{fig:amado}, we show the $p$ dependence of
$X(p,p_0,W)$ for $W=1$ and $p_0=\sqrt{4m(W+B)/3}$. 
The solid (dashed) curve represents the real (imaginary) part of the amplitude $X(p,p_0,W)$
extrapolated using the point method.
In the same figure, we also present the amplitude 
obtained by the spline interpolation method~\cite{Matsuyama:2006rp}
as a comparison.

The scattering amplitude $X(W)$ for the $\bar{K}NN$-$\pi YN$ system 
studied in this work is extrapolated from the amplitude $X(W+i\epsilon_l)$ 
at $\epsilon_l=10\times l$ (MeV) for $l=1,2,\dots,5$,
using Eqs.~(\ref{eq:cont_frac}) and~(\ref{eq:cont_coeff}).
\end{appendix}

\bibliographystyle{unsrt}

\end{document}